\documentclass{ws-ijmpe}
\usepackage[super,compress]{cite}

\begin{document}

\markboth{H.~Berrehrah, Th.~Steinert, W.~Cassing and E.~Bratkovskaya}{A dynamical quasiparticle approach for the Quark-Gluon-Plasma bulk and transport properties}

%%%%%%%%%%%%%%%%%%%%% Publisher's Area please ignore %%%%%%%%%%%%%%%
\catchline{}{}{}{}{}
%%%%%%%%%%%%%%%%%%%%%%%%%%%%%%%%%%%%%%%%%%%%%%%%%%%%%%%%%%%%%%%%%%%%

\title{A dynamical quasiparticle approach for the Quark-Gluon-Plasma bulk and transport properties}

\author{Hamza Berrehrah\footnote{berrehrah@fias.uni-frankfurt.de}}

\address{Frankfurt Institute for Advanced Studies, Johann Wolfgang Goethe Universit\"at, Ruth-Moufang-Strasse 1, 60438 Frankfurt am Main, Germany\\
Institute for Theoretical Physics, Johann Wolfgang Goethe Universit\"at, Max-von-Laue-Str. 1, 60438 Frankfurt am Main, Germany}

\author{Elena Bratkovskaya\footnote{E.Bratkovskaya@gsi.de}}

\address{Frankfurt Institute for Advanced Studies, Johann Wolfgang Goethe Universit\"at, Ruth-Moufang-Strasse 1, 60438 Frankfurt am Main, Germany\\
Institute for Theoretical Physics, Johann Wolfgang Goethe Universit\"at, Max-von-Laue-Str. 1, 60438 Frankfurt am Main, Germany \\
GSI Helmholtzzentrum f\"ur Schwerionenforschung GmbH,  64291
Darmstadt, Germany}

\author{Thorsten Steinert\footnote{thorsten.steinert@theo.physik.uni-giessen.de},
\ Wolfgang Cassing\footnote{wolfgang.cassing@theo.physik.uni-giessen.de}}

\address{Institut f\"ur Theoretische Physik, Universit\"at Gie\ss en, Heinrich-Buff-Ring 16, 35392 Gie\ss en, Germany}
%second\_author@group.com}

\maketitle

\begin{history}
\received{Day Month Year}
\revised{Day Month Year}
%\accepted{Day Month Year}
%\comby{(xxxxxxxxxx)}
\end{history}

\begin{abstract}
The properties of quantum-chromo dynamics (QCD) nowadays are
accessable by lattice QCD calculations at vanishing quark chemical
potential $\mu_q$=0 but often lack a transparent physical interpretation.
In this review we report about results from an extended dynamical
quasiparticle model (DQPM$^*$) in which the effective parton
propagators have a complex selfenergy that depends on the
temperature $T$ of the medium as well as on the chemical potential
$\mu_q$ and the parton three-momentum ${\boldsymbol p}$ with respect
to the medium at rest. It is demonstrated that this approach allows
for a good description of QCD thermodynamics with respect to the
entropy density, pressure etc. above the critical temperature $T_c
\approx$ 158 MeV. Furthermore, the quark susceptibility $\chi_q$ and
the quark number density $n_q$ are found to be reproduced
simultaneously at zero and finite quark chemical potential. The
shear and bulk viscosities $\eta, \zeta$, and the electric
conductivity $\sigma_e$  from the DQPM$^*$ also turn out in close
agreement with lattice results for $\mu_q$ =0. The DQPM$^*$,
furthermore, allows to evaluate the momentum $p$,  $T$ and  $\mu_q$
dependencies of the partonic degrees of freedom also for larger
$\mu_q$ which are mandatory for transport studies of heavy-ion
collisions in the regime 5 GeV $< \sqrt{s_{NN}} <$ 10 GeV. We
finally calculate  the charm quark diffusion coefficient $D_s$ --
evaluated from the differential cross sections of partons in the
medium for light and heavy quarks by employing the propagators and
couplings from the DQPM -- and compare to the available lattice
data. It is argued that the complete set of observables allows for a
transparent interpretation of the properties of hot QCD.
\end{abstract}


\ccode{PACS numbers:24.10.Jv, 02.70.Ns, 12.38.Mh, 24.85.+p}

\tableofcontents
%--------------------------------------------------------------------------------------------------------------------------------------------------------------------------------
\section{Introduction}
\label{Introduction}
%--------------------------------------------------------------------------------------------------------------------------------------------------------------------------------

The thermodynamic properties of the quark-gluon plasma (QGP)--as
produced in relativistic heavy-ion collisions--are rather well
determined within lattice QCD (lQCD) calculations at vanishing quark
chemical potential $\mu_q$
\cite{Borsanyi:2014rza,Borsanyi:2013bia,Borsanyi:2013cga,Borsanyi:2010bp,Borsanyi:2010cj,Soltz:2015ula}.
Whereas the results from different collaborations in the past have
led to different equations of state (EoS) of partonic matter even at
$\mu_q$ = 0 the present status can be considered as a consensus
(within error bars). Nevertheless, the physical interpretation of
the lattice 'data' remains a challenge since the EoS as well as
transport coefficients from lQCD indicate that the partonic system
cannot be viewed as a weakly interacting medium of quark,
antiquarks, and gluons. This holds especially true for temperatures
close to the critical temperature $T_c$ where the entropy density
$s(T)$ (and pressure $P(T)$) differ substantially from the Stefan
Boltzmann limit. The lQCD results on the EoS can conveniently be
interpreted within quasiparticle models with massive partons
\cite{Gorenstein:1995vm,Levai:1997yx,Peshier:1999ww,Peshier:2002ww,Bannur:2005wm,Srivastava:2010xa,Bluhm:2004xn,Plumari:2011mk}
that are fitted to the equation of state (EoS) from lQCD and also
allow for extrapolations to finite $\mu_q$, although with some
ambiguities. However, in these effective models the spectral
function of the degrees of freedom is taken as a $\delta$ - function
(on-shell limit) which implies that these partons in principle are
non-interacting. An extension of the simple quasiparticle model has
been proposed in Refs.
\cite{Peshier:2005pp,Cassing2007365,Cassing:2007nb,Cassing:2008nn}
where a finite width of the partonic spectral functions is
introduced, which corresponds to the interaction rate of the parton
in the medium at finite temperature $T$ and chemical potential
$\mu_q$. The latter can be directly employed for the calculation of
transport coefficients such as shear and bulk viscosities of the
partonic medium in the relaxation time approximation
\cite{Marty:2013ita} and be compared to corresponding correlators
from lQCD. An interpretation in terms of quasiparticles, however, is
constraint to effective propagators with a spectral width that is
substantially smaller than the dynamical pole mass.

Furthermore, at non-zero quark chemical potential $\mu_q \ne 0$, the
primary quantities of interest  are the ``pressure difference
$\Delta P$'', the quark number density $n_B$ and quark
susceptibility  $\chi_q$ since these quantities are available from
lQCD  \cite{Borsanyi:2012cr,Endrodi:2011gv}.  The quark number
susceptibilities are additional quantities to further quantify the
properties of the partonic degrees of freedom ( d.o.f.) especially
in the vicinity of the QCD phase transition or crossover
\cite{Bazavov:2009zn,Borsanyi:2010bp,Borsanyi:2010cj}. It turns out
that the standard quasiparticle models, that fit the partonic EoS,
severely underestimate the quark susceptibilities. Nevertheless, the
challenge of describing simultaneously both the lQCD pressure and
quark susceptibilities as well as transport coefficients is out of
reach in these models \cite{Bluhm:2004xn} which has been pointed out
in particular in Refs. \cite{Bluhm:2004xn,Plumari:2011mk}.
Especially the quark susceptibilities are very sensitive to the
quark masses used as inputs and solely determined by the quark
degrees of freedom. On the other hand both light quark and gluon
masses contribute to thermodynamic quantities like the entropy
density $s$ and pressure $P$. Therefore, reconciling all observables
from lQCD within a single effective model is a challenge.

Apart from the interactions in the light quark sector -- dominating
the partonic equation of state -- also the properties of heavy charm
quarks are of interest since their drag and diffusion controls the
elliptic flow $v_2$ of charm quarks as well as the suppression at
high transverse momentum \cite{ALICE:2012ab,Abelev:2013lca} in
relativistic heavy-ion reactions. Although the charm quarks can be
considered as reasonable quasiparticles -- with a pole mass that is
large compared to the spectral width -- the interactions with the
nonperturbative bulk partons are of interest and in particular the
transport coefficient ${\hat q}$ and the $c$-quark drag coefficient
as a function of $T$ and $\mu_q$.

In this review we will consider the QGP as a dynamical
quasi-particle medium of massive off-shell particles with partonic
propagators incorporating complex selfenergies which explicitly
depend on the three-momentum ${\boldsymbol p}$ with respect to the partonic
matter at rest in order to match perturbative QCD (pQCD) at high
momenta. We will show that within the extended dynamical
quasiparticle model -- denoted by DQPM$^*$ -- we reproduce the lQCD
equation of state at finite temperature $T$ and chemical potential
$\mu_q$. Moreover, we simultaneously describe the quark number
density and susceptibility $\chi_q$ from lQCD. In the same approach,
we also compute  the shear and bulk viscosities ($\eta$ and
$\zeta$), and the electric conductivity ($\sigma_e$)  of the QGP at
finite temperature $T$ and chemical potential $\mu_q$ in order to
probe some transport properties of the partonic medium in analogy to
the  studies in Refs.
\cite{Ozvenchuk:2012kh,Berrehrah:2013mua,Berrehrah:2014ysa,Berrehrah:2015ywa}.
The partonic spectral functions (or imaginary parts of the retarded
propagators) at finite temperature and chemical potential are
determined for these dynamical quasi-particles and the shear
viscosity $\eta$ and bulk viscosity $\zeta$ is computed within the
relaxation-time approximation (RTA) which provides similar results
as the Green-Kubo method employed in Refs.
\cite{Peshier:2005pp,Aarts:2004sd,Aarts:2002cc}.

The review is organized as follows:  We first present in Section
\ref{DQP-finiteTmu} the basic ingredients of the QGP d.o.f in terms
of their masses and widths, which are the essential ingredients in
their retarded propagators, as well as the running coupling
(squared) $g^2(T,\mu_q)$.  The gluon and fermion propagators
-- as given by the DQPM$^*$ at finite three-momentum ${\boldsymbol p}$,
temperature $T$ and quark chemical potential $\mu_q$ -- contain a
few parameters that are fixed in comparison to results from lQCD. As
a first application we will compute the lQCD pressure and
interaction measure in a partonic medium at finite $T$ and $\mu_q$
and compare to related results from lattice QCD. In Section 3  we
investigate the quark number density and susceptibility within the
DQPM$^*$ and compare to lQCD results for 2+1 flavors ($N_f = 3$). In
Section 4 we compute the QGP shear and bulk viscosities as well as
the electric conductivity and compare to lQCD results and other
theoretical studies.  Throughout Sections 2-4 we will point out the
importance of finite masses and widths of the light dynamical
quasiparticles, including their finite momentum, temperature and
$\mu_q$ dependencies. Section 5 is devoted to the dynamics and
transport properties of heavy charm quarks in the hot and dense
medium. To this aim we calculate the differential cross sections
between the light and heavy partons -- on the basis of the standard
DQPM couplings and propagators -- and evaluate their interaction
rates in the quasiparticle limit. Furthermore, we compute the
spatial diffusion coefficient and energy loss of the charm degrees
of freedom and compare to the available lQCD data. In Section 6,
finally, we summarize the main results and point out the future
applications of the DQPM$^*$.

%--------------------------------------------------------------------------------------------------------------------------------------------------------------------------------
\section{Parton properties in the DQPM$^*$}
\label{DQP-finiteTmu}
%--------------------------------------------------------------------------------------------------------------------------------------------------------------------------------
In the DQPM$^*$ the entropy density $s(T)$, the pressure $P(T)$ and
energy density $\epsilon(T)$ are calculated in a straight forward
manner by starting with the entropy density in the quasiparticle
limit from Baym
\cite{Cassing:2008nn,Cassing:2008sv,Vanderheyden:1998ph},{\setlength\arraycolsep{0pt}
\begin{eqnarray}
\label{sdqp} & & s^{dqp} = - d_g \!\int\!\!\frac{d \omega}{2 \pi}
\frac{d^3p}{(2 \pi)^3} \frac{\partial f_B}{\partial T} \left(
\Im\ln(-\Delta^{-1}) + \Im\Pi\,\Re\Delta \right)
\nonumber\\
& & {}    - d_q \!\int\!\!\frac{d \omega}{2 \pi} \frac{d^3p}{(2
\pi)^3} \frac{\partial f_F((\omega-\mu_q)/T)}{\partial T} \left(
\Im\ln(-S_q^{-1}) + \Im \Sigma_q\ \Re S_q \right)
 \!
\nonumber\\
& & {}    - d_{\bar q} \!\int\!\!\frac{d \omega}{2 \pi}
\frac{d^3p}{(2 \pi)^3} \frac{\partial
f_F((\omega+\mu_q)/T)}{\partial T} \left( \Im\ln(-S_{\bar q}^{-1}) +
\Im \Sigma_{\bar q}\ \Re S_{\bar q} \right) \!,
%\nonumber\\
\end{eqnarray}} where $f_B(\omega/T) = (\exp(\omega/T)-1)^{-1}$ and
$f_F((\omega-\mu_q)/T) = (\exp((\omega-\mu_q)/T)+1)^{-1}$ denote the
Bose and Fermi distribution functions, respectively, while $\Delta
=(P^2-\Pi)^{-1}$, $S_q = (P^2-\Sigma_q)^{-1}$ and $S_{\bar q} =
(P^2-\Sigma_{\bar q})^{-1}$ stand for the full (scalar)
quasiparticle propagators of gluons $g$, quarks $q$ and antiquarks
${\bar q}$.  In Eq. (\ref{sdqp}) $\Pi$ and $\Sigma = \Sigma_q
\approx \Sigma_{\bar q}$ denote the (retarded) quasiparticle
selfenergies. In principle, $\Pi$ as well as $\Delta$ are Lorentz
tensors and should be evaluated in a nonperturbative framework. The
DQPM$^*$ treats these degrees of freedom as independent scalar
fields with scalar selfenergies  which are assumed to be identical
for quarks and antiquarks. Note that one has to treat quarks and
antiquarks separately in Eq. (\ref{sdqp}) as their abundance differs
at finite quark chemical potential $\mu_q$. In Eq. (\ref{sdqp}) the
degeneracy for gluons is $d_g=2(N_c^2-1)$=16 while $d_q=d_{\bar
q}$=2$N_c N_f$=18 is the degeneracy for quarks and antiquarks with
three flavors. In practice one also has to differentiate between
$(u,d)$ and $s$ quarks due to their mass difference.

As a next step one writes the complex selfenergies as $\Pi({\boldsymbol q})
= M_g^2({\boldsymbol q})-2i \omega \gamma_g({\boldsymbol q})$ and $\Sigma_{q} ({\boldsymbol
q}) = M_{q} ({\boldsymbol q})^2 - 2 i \omega \gamma_{q} ({\boldsymbol q})$ with a
mass (squared) term $M^2$ and an interaction width $\gamma$, i.e.
the  retarded propagators ($\Delta, S_q$) read,
\begin{equation}
\label{propdqpm} G_R(\omega, {\boldsymbol q}) = \left(\omega^2 - {\boldsymbol q}^2 - M^2({\boldsymbol q}) + 2i
\gamma({\boldsymbol q}) \omega \right)^{-1},
\end{equation} and are analytic in the upper half plane in the
energy $\omega$ since the poles of $G_R$ are located in the lower half plane.
The imaginary part of $G_R$ (\ref{propdqpm}) then gives
the spectral function of the degree of freedom (except for a factor
$1/\pi$). In the standard DQPM
\cite{Cassing:2008nn,Bratkovskaya2011162,Peshier:2004ya} the masses
had been fixed in the spirit of the hard thermal loop (HTL) approach
with the masses being proportional to an effective coupling
$g(T/T_c)$ which has been enhanced in the infrared. In  the DQPM$^*$
the selfenergies depend additionally on the three-momentum ${\boldsymbol p}$
with respect to the medium at rest, while the dependence on the
temperature $T/T_c$ and chemical potential $\mu_q$ are very similar
to the standard DQPM
\cite{Cassing:2008nn,Bratkovskaya2011162,Peshier:2004ya}.

%%-------------------------------------------------------\
\subsection{Masses, widths and spectral functions of partons in DQPM$^*$}
%-------------------------------------------------------/

The functional forms for the parton masses and widths at finite
temperature $T$, quark chemical potential $\mu_q$ and momentum
$p=|{\boldsymbol p}|$ are assumed to be given by
{\setlength\arraycolsep{0pt}
\begin{eqnarray}
\label{equ:Sec2.1} & & M_g (T, \mu_q, p) =
\!\left(\frac{3}{2}\right) \!\! \Biggl[\frac{g^2
(T^{\star}/T_c(\mu_q))}{6} \Bigl[\bigl( N_c + \frac{N_f}{2} \bigr)
T^2  + \frac{N_c}{2} \sum_q \frac{\mu_q^2}{\pi^2} \Bigr]
\Biggr]^{1/2} \!\!\!\!\!\!\!\! \times h(\Lambda_g,p) + m_{\chi g}\;
,
\nonumber\\
& & {} M_{q,\bar{q}} (T, \mu_q, p)  = \Biggl[\frac{N_c^2-1}{8N_c}\,
g^2 (T^{\star}/T_c(\mu_q)) \Bigl[T^2 +  \frac{\mu_q^2}{\pi^2} \Bigr] \Biggr]^{1/2}
\!\!\! \times h(\Lambda_q,p) + m_{\chi q} \, ,
\nonumber\\
& & {} \gamma_g (T, \mu_q, p) = N_c\, \frac{g^2
(T^{\star}/T_c(\mu_q))}{8\pi} \ T \, \ln \left(\frac{2 c}{g^2
(T^{\star}/T_c(\mu_q))}+1.1 \right)^{3/4} \!\!\! \times h(\Lambda_g,p) \; ,
\nonumber\\
& & {} \gamma_{q,\bar{q}} (T, \mu_q, p) = \frac{N_c^2-1}{2N_c}\,
\frac{g^2 (T^{\star}/T_c(\mu_q))}{8\pi}\, T \ln \left(\frac{2 c}{g^2
(T^{\star}/T_c(\mu_q))}+1.1 \right)^{3/4} \!\!\! \times h(\Lambda_q,p) \ \,
,
\end{eqnarray}}
with the momentum-dependent function
\begin{equation} \label{hl} h(\Lambda,p)= \Bigl[\frac{1}{1 + \Lambda
(T_c(\mu_q)/T^{\star}) p^2} \Bigr]^{1/2},
\end{equation}
where $T^{\star 2} = T^2 +  \mu_q^2/\pi^2$ is the effective
temperature used to extend the DQPM$^*$ to finite $\mu_q$, while
$\Lambda_g (T_c(\mu_q)/T^{\star})$ = 5 $(T_c(\mu_q)/T^{\star})^2$
GeV$^{-2}$ and $\Lambda_q (T_c(\mu_q)/T^{\star}) =$ 12
$(T_c(\mu_q)/T^{\star})^2$ GeV$^{-2}$. Furthermore,  $m_{\chi g}
\approx 0.5$ GeV is the gluon condensate and $m_{\chi q}$ is the
light-quark chiral mass ($m_{\chi q} = 0.003$ GeV for $u$, $d$
quarks and $m_{\chi q} = 0.06$ GeV for $s$ quarks). Since the
effective quark masses in the QGP are large compared to the chiral
masses the latter can in practice be neglected. In Eq.
(\ref{equ:Sec2.1}) $m_{\chi g}$ ($m_{\chi q}$) gives the finite
gluon (light quark) mass in the limit $p \rightarrow 0$ and $T = 0$
or for $p \rightarrow \infty$. As mentioned above the quasiparticle
masses and widths (\ref{equ:Sec2.1}) are parametrized following hard
thermal loop (HTL) functional dependencies at finite temperature as
in the default DQPM \cite{Cassing:2008nn} in order to follow the
correct high temperature limit. The essentially new elements in
(\ref{equ:Sec2.1}) are the multiplicative factors $h(\Lambda,p)$
(\ref{hl}) specifying the momentum dependence of the masses and
widths with additional parameters $\Lambda_g$ and $\Lambda_q$ and
the additive terms $m_{\chi g}$ and $m_{\chi q}$. The
momentum-dependent factor $h(\Lambda,p)$ in the masses
(\ref{equ:Sec2.1}) is motivated by Dyson-Schwinger studies in the
vacuum \cite{Fischer:2006ub} and yields the limit of pQCD for $p
\rightarrow \infty$.

The effective gluon and quark masses are a function of $T^{\star}$
at finite $\mu_q$. Here we consider three light flavors $(q = u, d,
s)$ and assume all chemical potentials to be equal $(\mu_u = \mu_d = \mu_s
= \mu_q)$. Note that alternative settings are also possible to
comply with strangeness neutrality in heavy-ion collisions. The coupling (squared) $g^2$ in Eq. (\ref{equ:Sec2.1})
is the effective running coupling given as a function of $T/T_c$ at $\mu_q = 0$. A straight
forward extension of the DQPM$^*$ to finite $\mu_q$ is to consider the
coupling as a function of $T^{\star}/T_c (\mu_q)$ with a
$\mu_q$-dependent critical temperature $T_c (\mu_q)$,
{\setlength\arraycolsep{0pt}
\begin{eqnarray}
\label{equ:Sec2.2}
& & T_c (\mu_q) = T_c(\mu_q = 0) \sqrt{1 - \alpha \mu_q^2} \approx T_c(\mu_q = 0) \biggl(1 - \alpha/2 \ \mu_q^2 + \dots \biggr)
%\nonumber\\
%& & {}
\end{eqnarray}}
with $\alpha \approx 8.79$ GeV$^{-2}$. We recall that the expression
of $T_c(\mu_q)$ in Eq. (\ref{equ:Sec2.2}) is obtained by requiring a
constant energy density $\epsilon$ for the system at $T =
T_c(\mu_q)$ where $\epsilon$ at $T_c(\mu_q = 0) \approx 0.158$ GeV
is fixed by a lattice QCD calculation at $\mu_q = 0$. The
coefficient in front of the $\mu_q^2$-dependent part can be compared
to lQCD calculations at finite (but small) $\mu_B$ which gives
\cite{Bonati:2014rfa} {\setlength\arraycolsep{0pt}
\begin{eqnarray}
\label{equ:Sec2.3}
& & T_c (\mu_B) = T_c(\mu_B = 0) \biggl(1 - \kappa \left(\frac{\mu_B}{T_c (\mu_B = 0)}\right)^2 + \dots \biggr)
\end{eqnarray}}
with $\kappa = 0.013(2)$. Rewriting (\ref{equ:Sec2.2}) in the form
(\ref{equ:Sec2.3}) and using $\mu_B \approx 3 \mu_q$ we get
$\kappa_{DQPM} \approx 0.0122$ which compares  well with the lQCD
result.

Using the pole masses and widths (\ref{equ:Sec2.1}), the spectral
functions for the partonic degrees of freedom are fully determined,
i.e. the imaginary parts of the retarded propagators. The real part
of the retarded propagators then follows from dispersion relations or directly from Eq. (2).
Since the retarded propagators show no poles in the upper complex
half plane in the energy $\omega$ the model propagators obey
micro-causality \cite{Rauber:2014mca}. The imaginary parts are of
 Lorentzian form and provide the spectral functions
\cite{Cassing2007365,Cassing:2007nb,Cassing:2009vt},
\begin{equation}
\label{equ:Sec2.4} \rho_i (\omega, \boldsymbol{p} ) =
\frac{\gamma_i({\boldsymbol p})}{\tilde{E}_i({\boldsymbol p})}
\biggl(\frac{1}{(\omega - \tilde{E}_i({\boldsymbol p}))^2 + \gamma_i^2({\boldsymbol p})} - \frac{1}{(\omega + \tilde{E}_i({\boldsymbol p}))^2 + \gamma_i^2({\boldsymbol p})} \biggr)
\end{equation}
with $ \tilde{E}_i^2 (\boldsymbol{p}) = \boldsymbol{p}^2 + M_i^2({\boldsymbol p}) -
\gamma_i^2({\boldsymbol p})$ for $i \in [g, q, \bar{q}]$. These spectral
functions (\ref{equ:Sec2.4}) are antisymmetric in $\omega$ and
normalized as
{\setlength\arraycolsep{-10pt}
\begin{eqnarray}
\label{equ:Sec2.5} & & \hspace{-0.6cm} \int_{- \infty}^{+ \infty}
\frac{d \omega}{2 \pi} \ \omega \ \rho_i (\omega, \boldsymbol{p}) =
\int_0^{+ \infty} \frac{d \omega}{2 \pi} \ 2 \omega \ \rho_i
(\omega, \boldsymbol{p}) = 1,
%\nonumber\\
%& & {}
\end{eqnarray}}
where $M_i(T,\mu_q,{\boldsymbol p})$, $\gamma_i(T,\mu_q,{\boldsymbol p})$ are the
particle pole mass and width at finite three momentum ${\boldsymbol p}$,
temperature $T$ and chemical potential $\mu_q$, respectively.

%--------------------------------------------\
\subsection{The running coupling in DQPM$^*$}
%--------------------------------------------/
In contrast to the previous DQPM studies in Refs.
\cite{Berrehrah:2013mua,Berrehrah:2014ysa,Berrehrah:2015ywa} we
report here a new solution for the determination of the effective
coupling  which is more flexible. The strategy to determine
$g^2(T/T_c)$ is the following: For every temperature $T$ we fit the
DQPM$^*$ entropy density (\ref{sdqp}) to the entropy density
$s^{lQCD}$ obtained by lQCD. In practice, it has been checked that
for a given value of $g^2$, the ratio $s(T,g^2)/T^3$ is almost
constant for different temperatures and identical to $g^2$ in case
of momentum-independent selfenergies, i.e. $\frac{\partial}{\partial
T} (s(T,g^2)/T^3) \approx 0$. Therefore the entropy density $s$ and
the dimensionless equation of state in the DQPM is a function of the
effective coupling only, i.e. $s(T,g^2)/s_{SB} (T) = f(g^2)$. The
functional form,
$$\displaystyle f(g^2) = \frac{1}{(1 + a_1  (g^2)^{a_2})^{a_3}} , $$
however, is also suited to describe $s^{lQCD}(T,g^2)/s_{SB}$ in case
of momentum-dependent selfenergies in the DQPM$^*$. By inverting
$f(g^2)$, one arrives at the following parametrization for $g^2$ as
a function of $s/s_{SB}$: {\setlength\arraycolsep{0pt}
\begin{eqnarray}
\label{equ:Sec2.6} g^2(s/s_{SB}, T)  \sim  \left( \frac{a}{T} + b
\right)  \Biggl(\left(\frac{s/s_{SB}}{d (T)} \right)^{v(T)} - 1
\Biggr)^{w(T)},
\end{eqnarray}}
with  $s_{SB} = 19/(9\pi^2 T^3)$. Since the entropy
density from lQCD has the proper high temperature limit, the
effective coupling $g^2$ also gives the correct asymptotics for $T
\rightarrow \infty$ and decreases as $g^2 \sim 1/\log(T^2)$. The
temperature-dependent parameters $v(T)$, $w(T)$ and $d (T)$ all have
the functional form: {\setlength\arraycolsep{0pt}
\begin{eqnarray}
\label{equ:Sec2.61}
f (T)  =  \frac{a}{(T^b + c)^d}  (T + e),
\end{eqnarray}}
where the parameters $a$, $b$, $c$, $d$ and $e$ are fixed once for
each function $v(T)$, $w(T)$ and $d (T)$.

Note that with the parametrization (\ref{equ:Sec2.6}) for
$g^2(s/s_{SB}, T)$ one can easily adapt to any equation of state and
therefore avoid a refitting of the coupling in case of new (or
improved) lattice data. However, the coupling (\ref{equ:Sec2.6}) is
valid only for a given number of quark flavors $N_f$ which is fixed
by the lQCD equation of state.

To obtain $g^2(T/T_c)$ from $g^2(s/s_{SB}, T)$, we proceed as
follows:
\begin{itemize}
\item Using the equation of state from the Wuppertal-Budapest collaboration
\cite{Borsanyi:2012cr},
which provide an analytical parametrization of the interaction
measure $I/T^4$,
{\setlength\arraycolsep{0pt}
\begin{eqnarray}
\label{equ:Sec2.7}
\frac{I(T)}{T^4} = \exp(-h_1/t - h_2/t^2). \Biggl(h_0 + \frac{f_0 (\tanh(f_1 . t + f_2) + 1)}{1 + g_1 . t + g_2 . t^2} \Biggr),
\end{eqnarray}}
 with $t = T/200$ MeV, $h_0 = 0.1396$, $h_1 = - 0.18$, $h_2 = 0.035$, $f_0 = 2.76$, $f_1 = 6.79$, $f_2 = - 5.29$, $g_1 = - 0.47$ and $g_2 = 1.04$,
% (or $f_0 = 2.76$, $f_1 = 6.79$, $f_2 = - 5.29$, $g_1 = - 0.47$ and $g_2 = 1.04.57$),
\item we calculate the pressure $P/T^4$ by
{\setlength\arraycolsep{0pt}
\begin{eqnarray}
\label{equ:Sec2.8}
\frac{P(T)}{T^4} = \int_0^T \frac{I(T_0)}{T_0^5} d T_0,
\end{eqnarray}}
\item and then the entropy density
{\setlength\arraycolsep{0pt}
\begin{eqnarray}
\label{equ:Sec2.9}
s/s_{SB} = \frac{I(T)/T^4 + 4 P/T^4}{19/(9\pi^2)}.
\end{eqnarray}}
\item Replacing $s/s_{SB}$ from Eq.(\ref{equ:Sec2.9}) in  Eq.(\ref{equ:Sec2.6}) we obtain $g^2(T/T_c)$.
\end{itemize}
The procedure outlined above yields $g^2(T/T_c)$ for $\mu_q=0$. For
finite $\mu_q$ we will make use of $g^2 (T/T_c) \rightarrow g^2
(T^{\star}/T_c(\mu_q))$, with the $\mu_q$-dependent critical
temperature $T_c (\mu_q)$ taken from Eq. (\ref{equ:Sec2.2}). The
running coupling (\ref{equ:Sec2.6})-(\ref{equ:Sec2.9}) permits for
an enhancement near $T_c$ as already introduced in Ref.
\cite{Peshier:2002ww}.

\begin{figure}[h!] %tbh!
\begin{center}
\begin{minipage}{13.8pc}
\includegraphics[width=13.8pc, height=14.5pc]{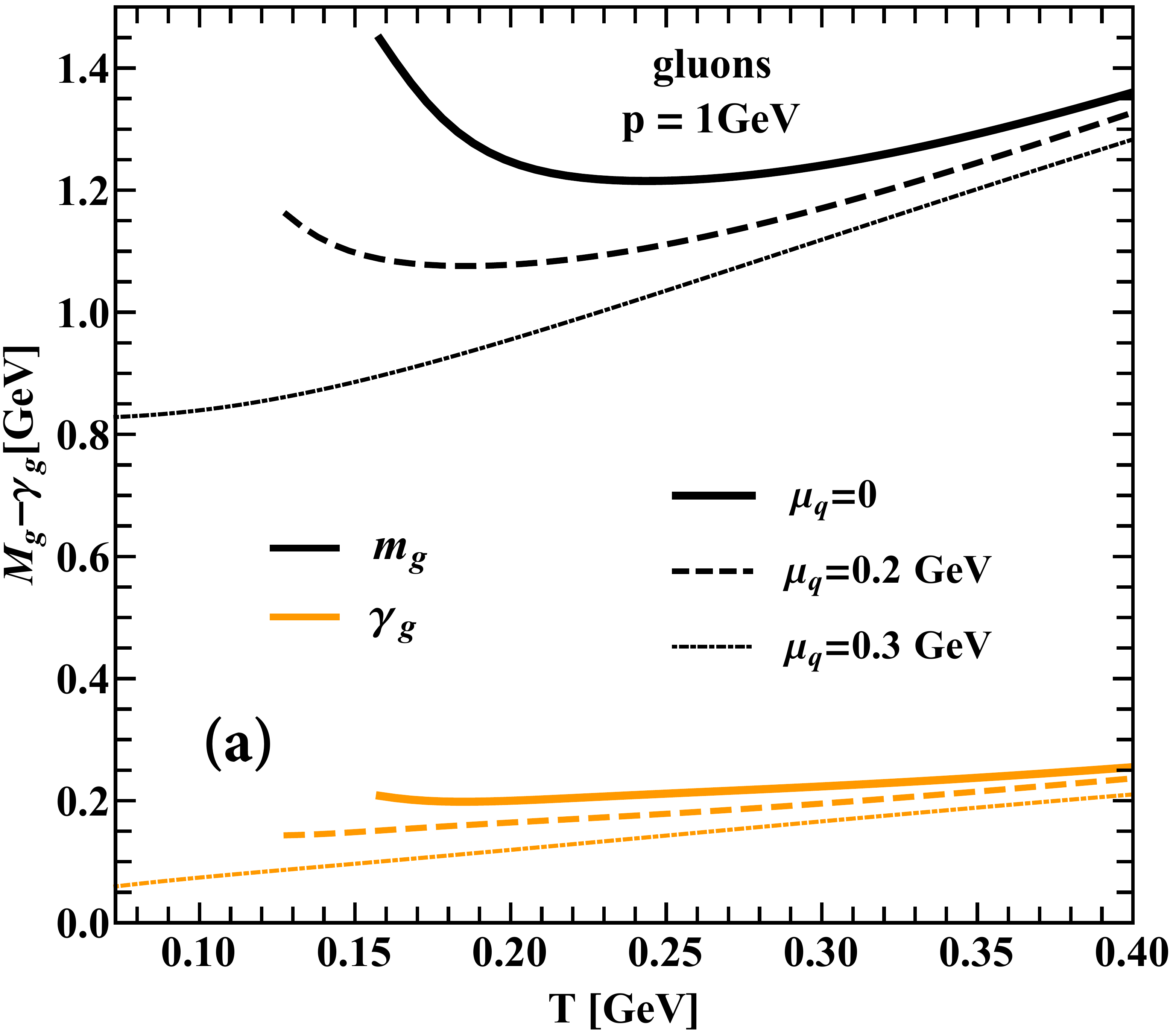}
\end{minipage} \hspace*{0.03cm}
\begin{minipage}{13.8pc}
\includegraphics[width=13.8pc, height=14.5pc]{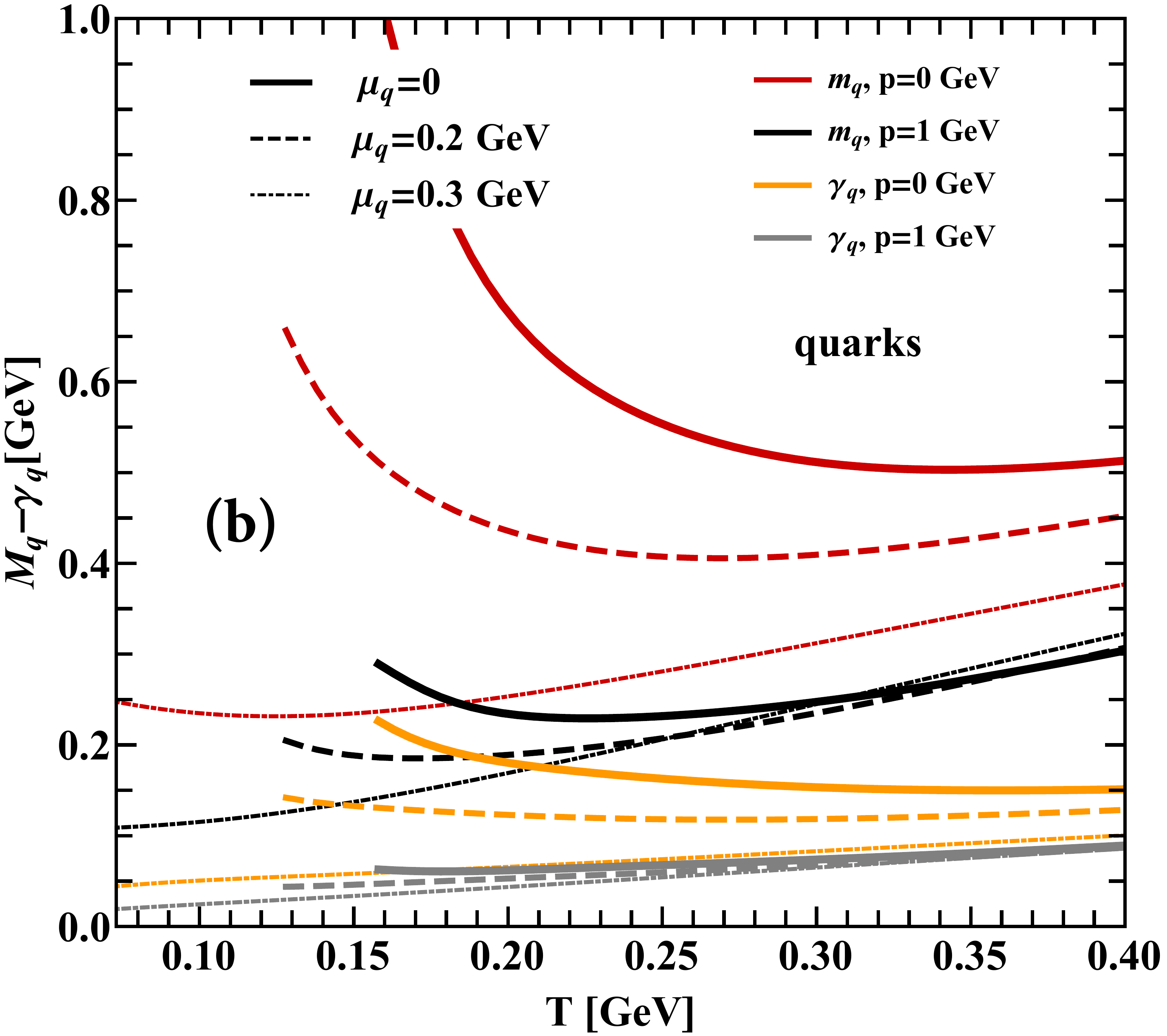}
\end{minipage} \hspace*{0.03cm}
\begin{minipage}{13.8pc}
\includegraphics[width=13.8pc, height=14.5pc]{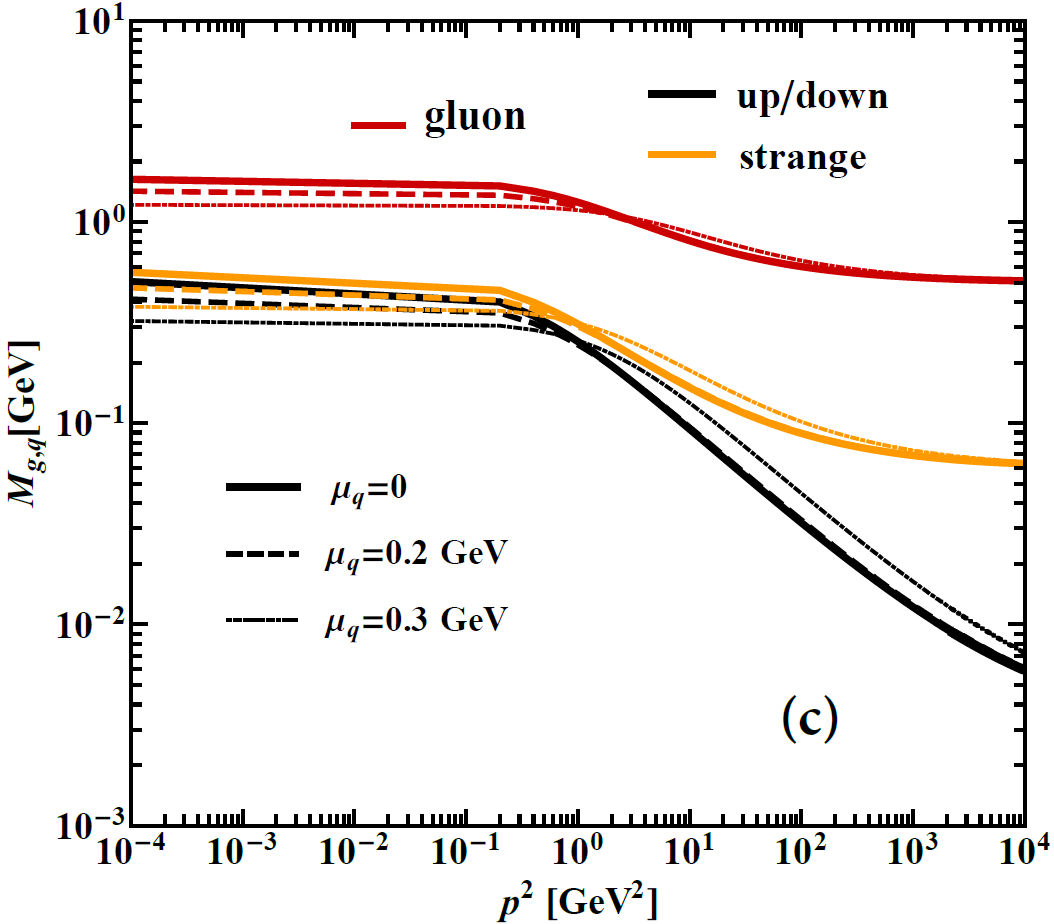}
\end{minipage}
\end{center}
%\vspace*{-0.5cm}
\caption{\emph{(Color online) The DQPM$^*$ gluon (a) and light quark (b) masses and widths given by (\ref{equ:Sec2.1})
using the coupling  (\ref{equ:Sec2.6})-(\ref{equ:Sec2.9}) for different quark chemical potentials as a
function of the temperature $T$. (c) Gluon and light quark masses as a function of the momentum squared for $T = 2 T_c$
and $\mu_q = 0, 0.2, 0.3$ GeV. The figures are taken from Ref. \cite{Berrehrah:2015vhe}.}}
\label{fig:mpDQPMIngred}
\end{figure}

Figs. \ref{fig:mpDQPMIngred} (a)-(b) show the gluon and light quark
masses and widths, respectively, at finite temperature and chemical
potential for a momentum $p = 1$ GeV/c. Furthermore, Fig.
\ref{fig:mpDQPMIngred} (c) shows the gluon and light quark masses as
a function of momentum (squared) $p^2$ at finite temperature $T = 2 T_c$ and different $\mu_q$. Note that for $p=0$ we obtain  higher
values of the gluon and light quark masses (as a function of $T$ and
$\mu_q$) since for finite  momenta the masses decrease (at a given
temperature and chemical potential), especially for the light quarks
as seen in Fig \ref{fig:mpDQPMIngred} (c). The extension $T/T_c \rightarrow T^{\star}/T_c(\mu_q)$ for finite $\mu_q$ in the
functional form for the strong coupling leads to lower values for
the parton masses and widths at finite $\mu_q$ as compared to
$\mu_q=0$ near $T_c (\mu_q)$.
%-(b) which give the 3D plots of the gluon and light quark masses as a function of $T$ and $p$ for $\mu_q =0$.
% As demonstrated in Fig.\ref{fig:mpDQPMIngred}, (a) the functional form for the strong coupling
%(\ref{equ:Sec3.3}) is in accordance with the lQCD calculations of Ref.\cite{Kaczmarek:2004gv} for the long range part of the $q-\bar{q}$ potential. We notice that $\alpha_s$ for $N_f=0$ is not present in Figs \ref{fig:mpDQPMIngred}-(a) because the coupling constant (\ref{equ:Sec6.1})-(\ref{equ:Sec6.4}) is determined for $N_f=3$ only.
\begin{figure}[h!]
\begin{center}
\begin{minipage}{12pc}
\includegraphics[width=12pc, height=11.5pc]{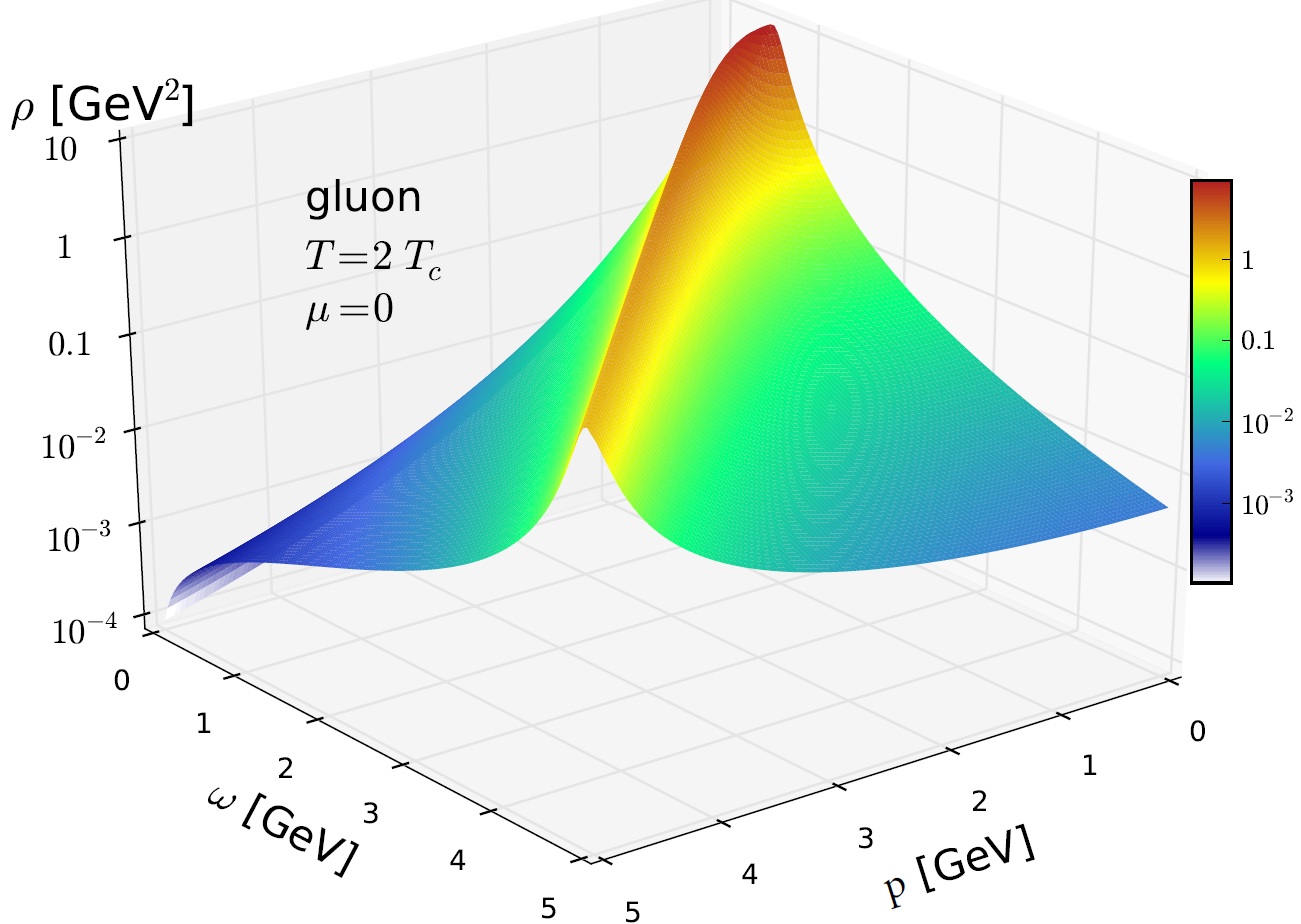}
\end{minipage} \hspace{0.1cm}
\begin{minipage}{12pc}
\includegraphics[width=12pc, height=11.5pc]{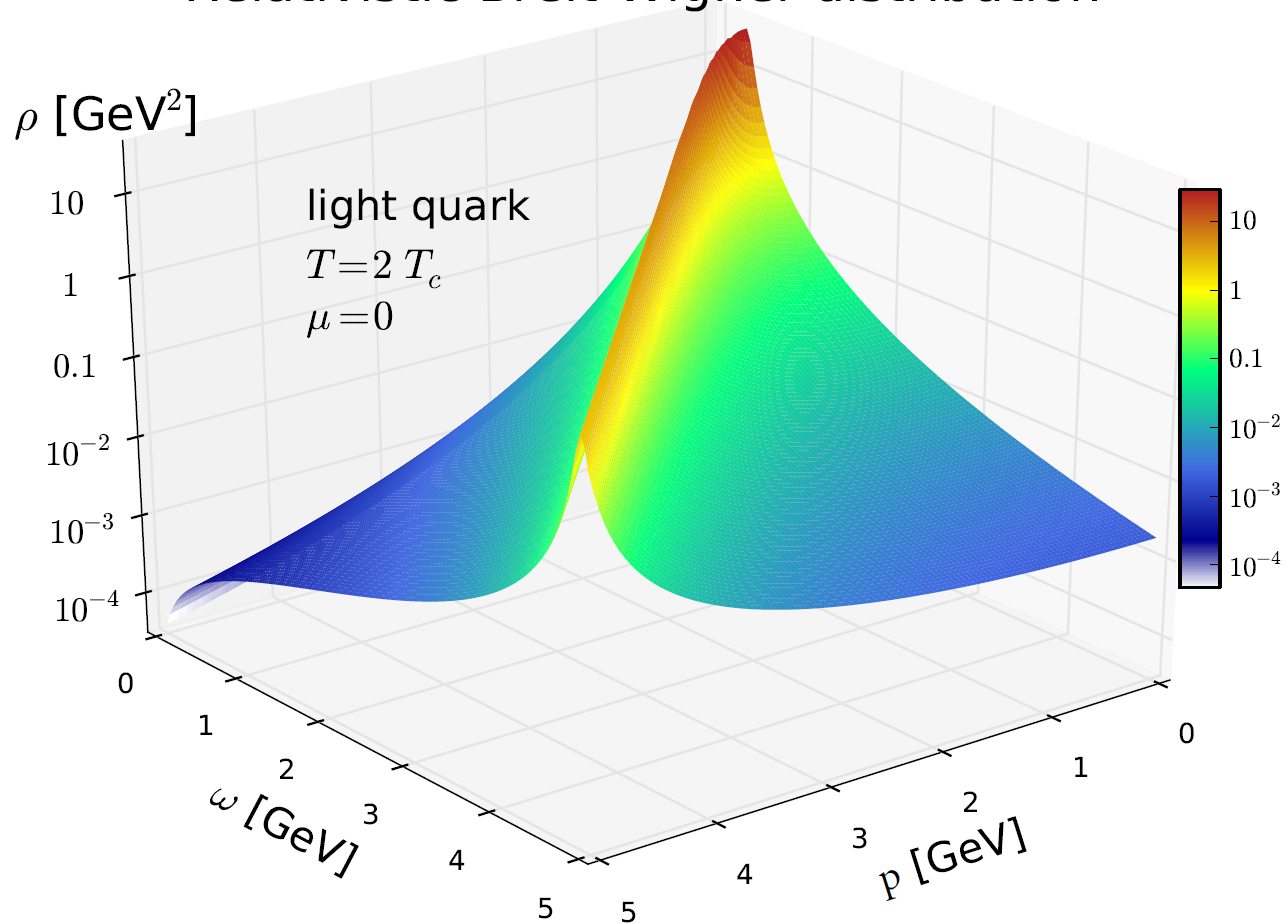}
\end{minipage} \hspace*{0.03cm}
\end{center}
\caption{\emph{(Color online) The DQPM$^*$ gluon (l.h.s.) and light
quark (r.h.s.) spectral functions at $T$ = 200 MeV as a function of
momentum $p$ and energy $\omega$.}} \label{fig2x}
\end{figure}

An illustration of the actual spectral functions (in $\omega$ and
momentum $p$) is given in Fig. \ref{fig2x} for a 'gluon' (l.h.s.)
and a light 'quark' (r.h.s.) at temperature $T$ = 200 MeV for
$\mu_q$=0.

%--------------------------------------------------------------------------------------------------------------------------------------------------------------------------------
\subsection{Thermodynamics of the QGP from DQPM$^*$}
\label{QGPthermo}
%--------------------------------------------------------------------------------------------------------------------------------------------------------------------------------

The expressions for the equation of state (energy density
$\epsilon$, entropy density $s$ and pressure $P$) of strongly
interacting matter have been given for finite temperature and
chemical potential in Ref. \cite{Marty:2013ita} for on-shell partons
and in Ref. \cite{Cassing:2008nn} for the case of off-shell partons
using the relations based on the stress-energy tensor $T^{\mu \nu}$.
We recall that the approach for calculating the equation of state in
the DQPM$^*$ is based on thermodynamic relations (see below). The
procedure is as follows: One starts from the evaluation of the
entropy density $s$ from (\ref{sdqp}) employing the masses and
widths obtained from the expressions in Section 2.1. Then using the
thermodynamic relation $s = ({\partial P}/{\partial T})_{\mu_q}$
(for a fixed quark chemical potential $\mu_q$) one obtains the
pressure $P$ by integration of the entropy density $s$ over $T$ while the energy density
$\epsilon$ can be gained using the relation,
\begin{equation}  \epsilon (T, \mu_B)= T s (T, \mu_B) - P (T, \mu_B) + \mu_B
n_B (T, \mu_B) , \end{equation} where $n_B$ is the net baryon
density.

\begin{figure}[h!] %tbh!
\begin{center}
\begin{minipage}{16pc}
\includegraphics[width=13pc, height=13pc]{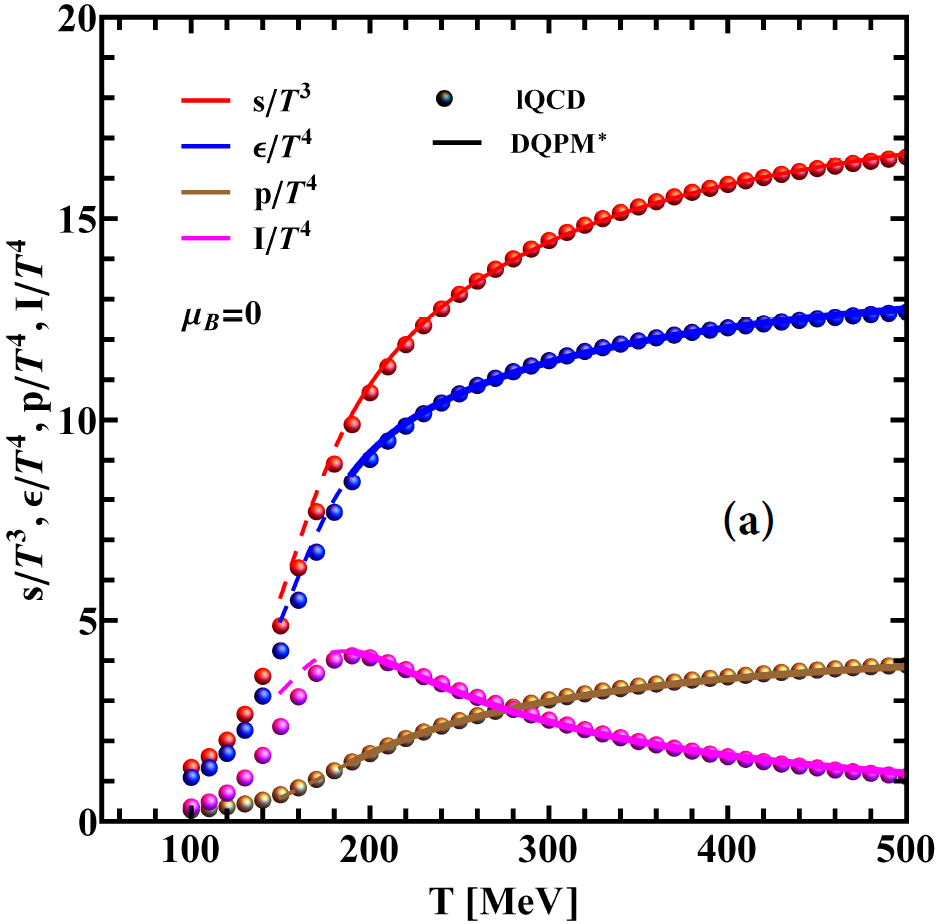}
\end{minipage}\hspace*{0.1cm}
\begin{minipage}{16pc}
\includegraphics[width=13pc, height=13pc]{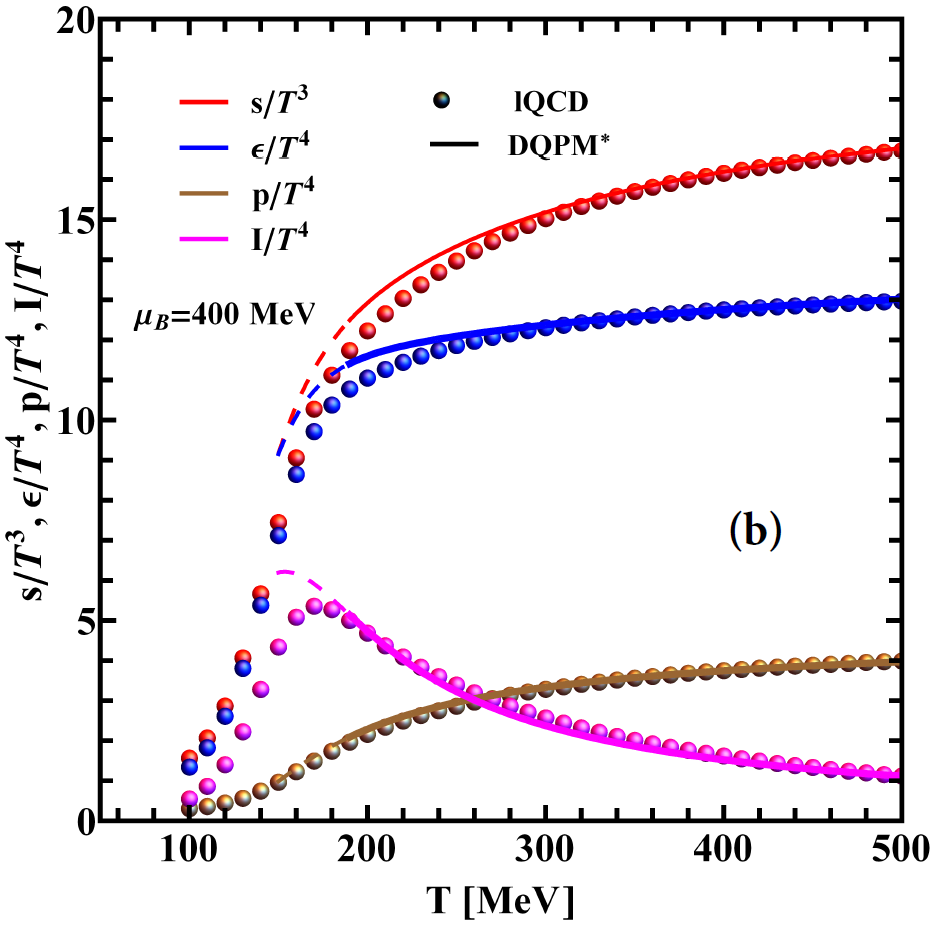}
\end{minipage}
\end{center}
\caption{\emph{(Color online) Scaled energy density $\epsilon$, entropy
density $s$, pressure $P$ and trace anomaly $(I = \epsilon -
3 P)$ as a function of temperature $T$ at $\mu_B = 0$ (a) and at
$\mu_B=400$ MeV (b) from DQPM$^*$ compared to lQCD data from Ref.
\cite{Borsanyi:2012cr}. The figures are taken from Ref. \cite{Berrehrah:2015vhe}.}} \label{fig:EoSTmu}
\end{figure}

The energy density $\epsilon$, entropy density $s$, pressure $P$ and
the interaction measure $[ I (T, \mu_q) = \epsilon (T, \mu_q) - 3 P
(T, \mu_q) ]$ --known in lQCD as the trace anomaly-- in the DQPM$^*$
are shown in Fig. \ref{fig:EoSTmu} (a), (b) as a function of
temperature $T$  for two values of the baryon chemical potential
$\mu_B = 0$ and $\mu_B = 400$ MeV, respectively (where $\mu_B = 3
\mu_q$ in our study). We, furthermore, compare our results with
lattice calculations from Ref. \cite{Borsanyi:2012cr} and notice
that our results are in a very good agreement with the lattice data
for $\mu_B = 0$ (a) and in case of $\mu_B = $ 400 MeV (b) for
temperatures larger than $1.2 \ T_c (\mu_q)$. In the latter case we
observe (for temperatures just above $T_c (\mu)$) some deviations
which are expected to result from additional hadronic degrees of
freedom in the crossover region.  The small excess in quarks can be
seen also in the net baryon density $n_B$, as we will show below.

At finite baryon chemical potential, i.e. $\mu_B = 400$ MeV, the
maximum of the trace anomaly  is shifted towards  lower
temperatures. We notice also the proper scaling of our DQPM$^*$
description of QGP thermodynamics, when moving from zero to finite
quark chemical potential (cf. Fig.\ref{fig:EoSTmu} (a) and (b)).

%--------------------------------------------------------------------------------------------------------------------------------------------------------------------------------
\section{Quark number density and susceptibility from DQPM$^*$}
\label{qChi}
%--------------------------------------------------------------------------------------------------------------------------------------------------------------------------------

\subsection{Baryon number density in the DQPM$^*$}
The equation of state for vanishing chemical potential $\mu_q$=0 is defined
solely by the entropy density $s$; for finite chemical potential
$\mu_q \neq 0$ one has to include the particle density $n$. In the DQPM$^*$
the quark density $n^{dqp}$ in the quasiparticle limit is defined in
analogy to the entropy density (\ref{sdqp}) as
\cite{Blaizot:2000fc},{\setlength\arraycolsep{0pt}
    \begin{eqnarray}
    \label{ndqp}
    n^{dqp} = & & \ - d_q \!\int\!\!\frac{d \omega}{2 \pi} \frac{d^3p}{(2 \pi)^3} \frac{\partial f_F((\omega-\mu_q)/T)}{\partial \mu_q} \left( \Im\ln(-S_q^{-1}) + \Im \Sigma_q\ \Re S_q \right)
    \!
    \nonumber\\
    & & {}    - d_{\bar q} \!\int\!\!\frac{d \omega}{2 \pi} \frac{d^3p}{(2 \pi)^3}
    \frac{\partial f_F((\omega+\mu_q)/T)}{\partial \mu_q} \left( \Im\ln(-S_{\bar q}^{-1}) + \Im \Sigma_{\bar q}\ \Re S_{\bar q} \right) \! .
    %\nonumber\\
    \end{eqnarray}}
In case of the Lorentzian spectral function (7) the density
$n^{dqp}$ in Eq. (\ref{ndqp}) can be split into the following two
terms $n_q^{(0)}$ and $\Delta n_q$ as: {\setlength\arraycolsep{-1pt}
\begin{eqnarray}
\label{equ:Sec4.3bis} & &  n_q^{(0)} = d_q \int \frac{d^3 p}{(2
\pi)^3} \ f_q^{(0)} - d_{\bar{q}} \int \frac{d^3 p}{(2 \pi)^3} \
f_{\bar{q}}^{(0)},
%\nonumber\\
%& & {}
\end{eqnarray}}
{\setlength\arraycolsep{0pt}
\begin{eqnarray}
\label{equ:Sec4.3tr} & & \Delta n_q = \int \frac{d \omega}{(2 \pi)}
\frac{d^3 p}{(2 \pi)^3} \frac{\partial f_q ((\omega -
\mu_q)/T)}{\partial \mu_q} \ \xi(\omega,p)
\nonumber\\
& & {} + \int \frac{d \omega}{(2 \pi)} \frac{d^3 p}{(2 \pi)^3}
\frac{\partial f_{\bar{q}} ((\omega + \mu_q)/T)}{\partial \mu_q} \
\xi(\omega,p) ,
\end{eqnarray}}
with \begin{equation} \label{equ:Sec4.3qr} \xi(\omega,p) = \Biggl(2 \gamma \omega
\frac{\omega^2 - \textbf{p}^2 - M^2}{(\omega^2 - \textbf{p}^2 -
M^2)^2 + 4 \gamma^2 \omega^2} - \arctan \left(\frac{2 \gamma
\omega}{\omega^2 - \textbf{p}^2 - M^2} \right) \Biggr)
\end{equation}
where $f_q^{(0)} = (\exp ((\sqrt{p^2 + M^2} - \mu_q)/T)+1)^{-1}$,
$f_{\bar{q}}^{(0)} = (\exp ((\sqrt{p^2 + M^2} + \mu_q)/T)+1)^{-1}$
denote again the Fermi distribution functions for the on-shell quark
and anti-quark, with $M$ corresponding to the pole mass.

Finally, note that the quark number density (\ref{ndqp}) follows
from the same potential as the entropy density
\cite{Vanderheyden:1998ph} which ensures that it fulfills the
thermodynamic relation $n=(\partial P/\partial \mu_q)_T$ (for fixed
temperature T). To be fully thermodynamically consistent the entropy
and the particle density have to satisfy the Maxwell relation
$(\partial n/\partial T)_{\mu_q}=(\partial s/\partial \mu_q)_T$.
This provides further constraints on the effective coupling
$g^2(T,\mu_q)$ at finite chemical potential which we neglect in the
current approach. Nevertheless, it was checked that the violation of
the latter Maxwell relation is generally small and most pronounced around
$T_c$. We note, however, that when extending the approach to even
larger chemical potentials the full thermodynamic consistency has to
be taken into account. The baryon number density $n_B$, finally, is
related to the quark number density by the simple relation
$n_B=n^{dqp}/3$.

%--------------------------------------------
\subsection{Susceptibilities in the DQPM$^*$}
%---------------

From the densities $n_B$ one may obtain other thermodynamic
quantities like the pressure difference $\Delta P$ and the quark
susceptibilities $\chi_q$, which can be confronted with lattice data
for $N_f=2$ from Alton \emph{et al}.
\cite{Allton:2003vx,Allton:2005gk} and for $N_f = 3$ from Borsanyi
\emph{et al}.\cite{Borsanyi:2012cr}. We recall that the quark-number
susceptibility measures the static response of the quark number
density to an infinitesimal variation of the quark chemical
potential. From Eqs. (\ref{equ:Sec4.3bis})-(\ref{equ:Sec4.3qr}) we calculate
$\Delta P$ and $\chi_q$ as {\setlength\arraycolsep{0pt}
\begin{eqnarray}
\label{equ:Sec4.4}
& & \Delta P (T, \mu_B) \equiv  P (T, \mu_B) - P (T,0) = \int_0^{\mu_B} n_B \ d \mu_B \ ;
%\nonumber\\
%& & {}
\end{eqnarray}}
{\setlength\arraycolsep{0pt}
\begin{eqnarray}
\label{equ:Sec4.5} & & \chi_q (T) = \frac{\partial n_q}{\partial
\mu_q} \biggl|_{\mu_q=0}; \hspace*{0.5cm} \chi_q (T, \mu_q) =
\frac{1}{9} \frac{\partial n_B}{\partial \mu_B}\ .
%\nonumber\\
%& & {}
\end{eqnarray}}
Furthermore, for small $\mu_q$ a Taylor expansion of the pressure in
$\mu_q/T$ can be performed which gives {\setlength\arraycolsep{0pt}
\begin{eqnarray}
\label{equ:Sec4.6}
& & \frac{P (T, \mu_q)}{T^4} = \sum_{n=0}^{\infty} c_n (T) \left(\frac{\mu_q}{T} \right)^n,
\hspace*{0.5cm}  c_n (T) = \frac{1}{n!} \frac{\partial^n (P(T, \mu_q)/T^4)}{\partial (\mu_q/T)^n}\Biggr|_{\mu_q = 0},
%\nonumber\\
%& & {}
\end{eqnarray}}
where $c_n(T)$ is vanishing for odd $n$ and $c_0 (T)$ is given by
$c_0 (T) = P (T, \mu_q = 0)$. As shown above the  DQPM$^*$ compares
well with lattice QCD results for $c_0(T)$. Since $\chi_q$ at finite
$\mu_q$ is related to the pressure by $$\displaystyle \chi_q (T,
\mu_q)/T^2 =
\partial^2 (P/T^4)/\partial^2 (\mu_q/T) ,$$ one can define the
susceptibility $\chi_2^{i j}$ at vanishing quark chemical potential
as \cite{Borsanyi:2012cr} {\setlength\arraycolsep{0pt}
\begin{eqnarray}
\label{equ:Sec4.7} & & \frac{P(T,{\mu_i})}{T^4} = \frac{P(T,
{0})}{T^4} + \frac{1}{2} \sum_{i, j} \frac{\mu_i \mu_j}{T^2}
\chi_2^{i j}, \hspace*{0.1cm} \textrm{with} \ \ \chi_2^{i j} =
\frac{1}{T^2} \frac{\partial n_j (T, {\mu_i})}{\partial \mu_i}
\Biggr|_{\mu_i = \mu_j = 0} ,
%\nonumber\\
%& & {}
\end{eqnarray}}
which in case of 3 flavors  with $\mu_u = \mu_d = \mu_s$ becomes
{\setlength\arraycolsep{0pt}
\begin{eqnarray}
\label{equ:Sec4.8} & & \chi_2 (T) = \frac{1}{9} \ \frac{1}{T^2}
\frac{\partial n_q (T, \mu_q)}{\partial \mu_q} \Biggr|_{\mu_q = 0} =
\frac{1}{9} \ \frac{\chi_q (T)}{T^2}.
%\nonumber\\
%& & {}
\end{eqnarray}}
We recall again that the susceptibilities are the central quantities
in lQCD calculations for nonzero $\mu_q$.

\subsection{$n_B$ and $\chi_q$: DQPM$^*$ \emph{vs} lQCD} %---------------

\begin{figure}[h!] %tbh!
\begin{center}
\begin{minipage}{14.5pc}
\includegraphics[width=14.5pc, height=15pc]{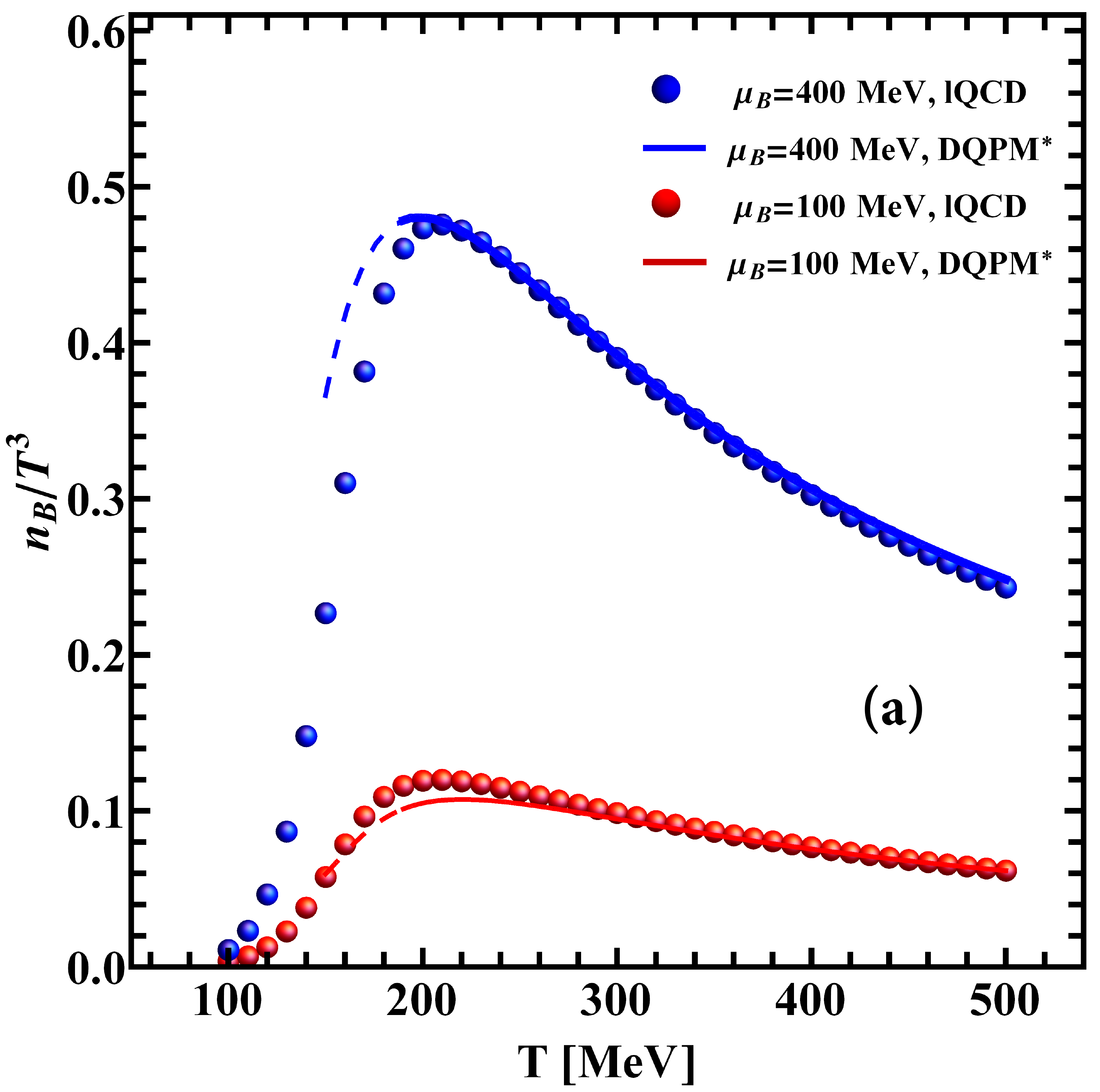}
\end{minipage} \hspace*{0.1cm}
\begin{minipage}{14.5pc}
\includegraphics[width=14.5pc, height=15pc]{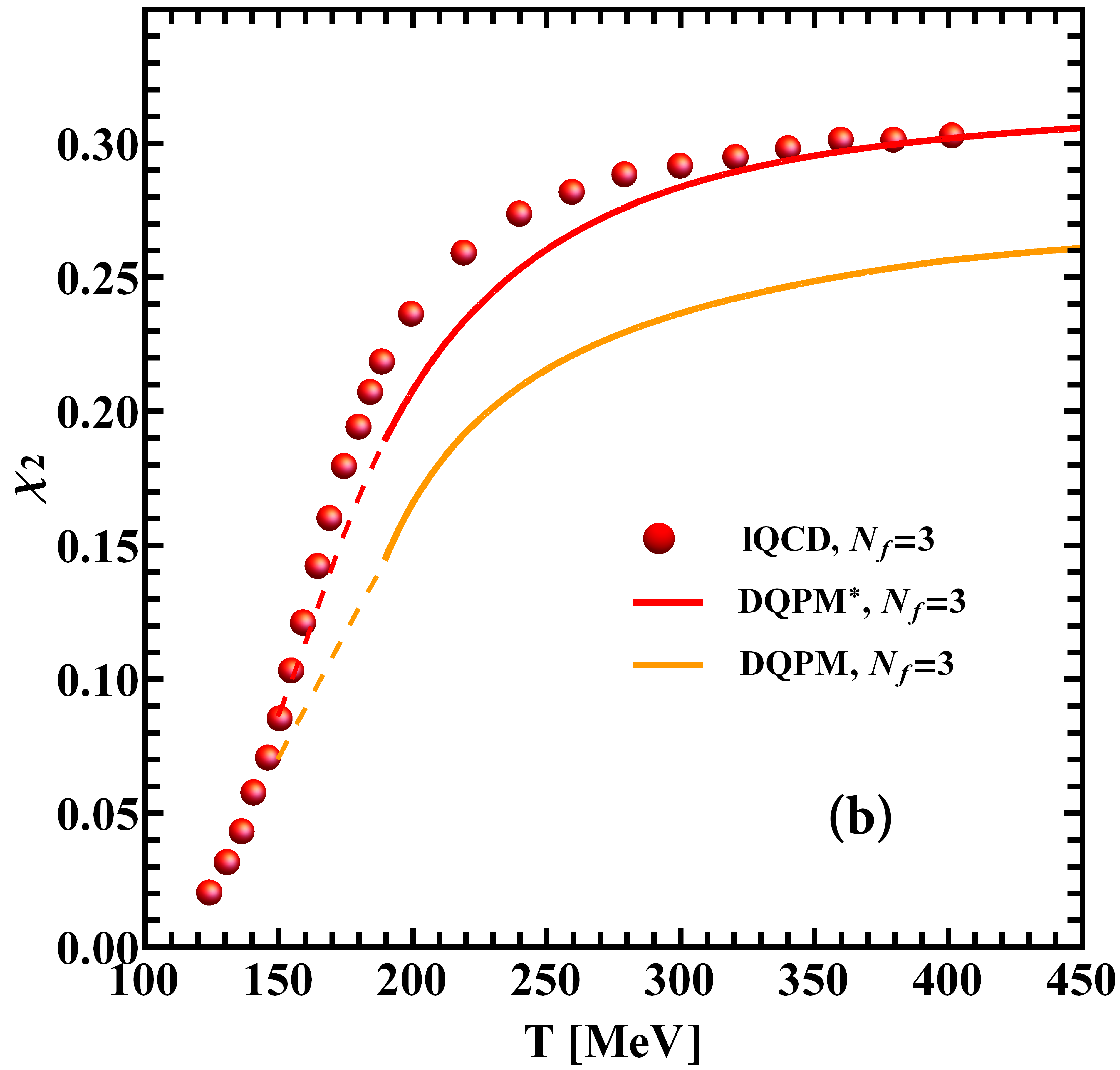}
\end{minipage}
\end{center}
\caption{\emph{(Color online) (a) The baryon number density
$n_B/T^3$ from DQPM$^*$ as compared to lattice data from Ref.
\cite{Borsanyi:2012cr} for $N_f = 3$ for a quark chemical potential
$\mu_q = 0$. (b) The susceptibility $\chi_2$ from DQPM$^*$ as
compared to lattice data from Ref.\cite{Borsanyi:2012cr} for $N_f = 3$
and $\mu_q=0$ using Eq. (\ref{equ:Sec4.8}). The lower (orange) line
gives the result from the conventional DQPM, i.e. with momentum
independent masses. The figures are taken from Ref. \cite{Berrehrah:2015vhe}.}} \label{fig:nBchi2Nf3}
\end{figure}

Using the masses and widths (\ref{equ:Sec2.1}) and the running
coupling (\ref{equ:Sec2.6})-(\ref{equ:Sec2.8}), we calculate the
baryon number density $n_B$
(\ref{equ:Sec4.3bis})-(\ref{equ:Sec4.3tr}) and quark susceptibility
$\chi_2$ including the finite width of the parton spectral
functions. The results for $n_B$ and $\chi_2$ for $N_f = 3$ are
given in Fig. \ref{fig:nBchi2Nf3} (a) and (b), respectively. The
comparison with the lattice data from Ref. \cite{Borsanyi:2012cr} is
rather good which is essentially due to an extra contribution
arising from the momentum dependence of the DQPM$^*$ quasiparticles
masses and widths. Such a momentum dependence in $m_{q,\bar{q},g}$
and $\gamma_{q,\bar{q},g}$ decreases the 'thermal average' of light
quark and gluon masses which improves the description of lQCD
results for the susceptibilities. For comparison we also show the
result for $\chi_q$ from the conventional DQPM, i.e. with momentum
independent masses, which substantially underestimates the lattice
data. The small difference between lQCD and DQPM$^*$ for $n_B$ and
$\chi_2$ close to $T_c$ is related to a possible excess of light
quarks and antiquarks  which should combine to hadrons in the
crossover region. We recall that the DQPM$^*$ describes only the QGP
phase and deals with dynamical quarks and gluons solely.

Finally, we emphasize the challenge to describe simultaneously the
entropy $s$ and pressure $P$ on one side and $n_B$ and $\chi_2$ on
the other side. Indeed, increasing the light quark mass and width
helps to improve the description of $s$ and $P$ (for $\mu_B$ = 400
MeV), but this leads to a considerable decrease in $n_B$ and
$\chi_2$. In other words, lighter quarks are favorable to improve
the agreement with lQCD data on $n_B$ and $\chi_2$, however, this
leads to an increase of $s$ and $P$, which can be only partially
counterbalanced by an increasing gluon mass and width (which do not
enter $n_b$ and $\chi_2$).

%--------------------------------------------------------------------------------------------------------------------------------------------------------------------------------
\section{Transport properties of the hot QGP from DQPM$^*$}
\subsection{Shear and bulk viscosities}
\label{etaOvs}
%--------------------------------------------------------------------------------------------------------------------------------------------------------------------------------

In this Section we focus on the transport coefficients of the QGP using the
relaxation time approximation (RTA). In the dilute gas approximation
the relaxation time $\tau_i$ of the particle $i$ is obtained for on-
or off-shell quasi-particles by means of the partonic scattering
cross sections, where the $qq$, $q\bar{q}$, $qg$ and $gg$ elastic
scattering processes as well as some inelastic processes involving
chemical equilibration, such as $gg \rightarrow q \bar{q}$ are
included in the computation of $\tau_i$ \cite{Berrehrah:2014ysa}.
For the DQPM$^*$ approach  we do not need the explicit cross
sections since the inherent quasi-particle width $\gamma_i (T,\mu_q,
p)$ directly provides the total interaction rate
\cite{Cassing:2008nn}. To this end we only have to evaluate the
average of the momentum dependent widths $\gamma_g (T,\mu_q, p)$ and
$\gamma_q (T,\mu_q, p)$ over the thermal distributions at fixed $T$
and $\mu_q$, i.e. $\bar \gamma_g (T,\mu_q)$ and $\bar \gamma_q
(T,\mu_q)$.

The shear viscosity $\eta (T,\mu_q)$ is defined in the dilute gas
approximation for the case of the DQPM$^*$ off-shell particles by
\cite{Chakraborty:2010fr,Berrehrah:2014ysa}
{\setlength\arraycolsep{0pt}
\begin{eqnarray}
\label{equ:Sec5.1}
\eta (T,\mu_q) = & & \ \frac{1}{15 T} d_g \!\!\int\!\! \frac{d^3 p}{(2\pi)^3} \ \!\!\int\!\!
\frac{d\omega}{2 \pi} \omega \ \bar \tau_g (T, \mu_q) \ f_g(\omega/T) \times \rho_{g}
(\omega, {\boldsymbol p}) \frac{{\boldsymbol p}^4}{\omega^2} \Theta(P^2)
\nonumber\\
& & {} + \frac{1}{15 T} \frac{d_q}{6} \!\!\int\!\! \frac{d^3
p}{(2\pi)^3} \ \!\!\int\!\! \frac{d \omega}{2 \pi} \omega \  \Bigg[
\sum_q^{u,d,s} \bar \tau_q (T, \mu_q) f_q((\omega-\mu_q)/T)
\rho_{q}(\omega,{\boldsymbol p})
\nonumber\\
& & {} \hspace*{0.6cm} + \sum_{\bar q}^{\bar u,\bar d,\bar s}
\bar \tau_{\bar q} (T, \mu_q) f_{\bar q}((\omega+\mu_q)/T) \ \rho_{\bar
q}(\omega,{\boldsymbol p}) \  \Bigg] \frac{{\boldsymbol p}^4}{\omega^2} \Theta(P^2),
\end{eqnarray}}
where ${\boldsymbol p}$ is the three-momentum and $P^2$ the invariant mass
squared. The functions $\rho_g, \rho_q, \rho_{\bar q}$ stand for the
gluon, quark and antiquark spectral functions, respectively, and
$f_q$ $(f_{\bar q})$ stand for the equilibrium distribution
functions for particle and antiparticle. The medium-dependent
relaxation times $\bar \tau_{q,g} (T, \mu_q)$ in (\ref{equ:Sec5.1}) are
given in the DQPM$^*$ by:

{\setlength\arraycolsep{-1pt}
\begin{eqnarray}
\label{equ:Sec5.2} \bar \tau_{q,g} (T, \mu_q) = (\bar \gamma_{q,g})^{-1}
(T,\mu_q),
\end{eqnarray}}
with:
{\setlength\arraycolsep{-1pt}
\begin{eqnarray}
\label{equ:Sec5.3} & &  \bar \gamma_{q,g} (T, \mu_q) = \displaystyle \langle\gamma_{q,g} (T, \mu_q, p)\rangle_p
\nonumber\\
& & {} = \left(n_{q,g}^{\textrm{off}} (T,\mu_q)\right)^{-1}\!\!\! \times \!\! \displaystyle \int \frac{d^3p}{(2 \pi)^3} \frac{d \omega}{(2 \pi)} \ \omega \ \gamma_{q,g} (T, \mu_q, p) \rho_f (\omega) f_{q,g} (\omega, T, \mu_q) \Theta(P^2),
\end{eqnarray}}
where $$n_{f,g}^{\textrm{off}} (T,\mu_q) = \displaystyle \int
\frac{d^3p}{(2 \pi)^3} \ \frac{d \omega}{(2 \pi)} \ \omega \ \rho_f
(\omega) \ f_{f,g} (\omega, T, \mu_q) \ \Theta(P^2),$$ denotes the
off-shell density of quarks, antiquarks or gluons. We note in
passing that the shear viscosity $\eta$ can also be computed using
the stress-energy tensor and the Green-Kubo formalism
\cite{Ozvenchuk:2012kh}. However, explicit comparisons of both
methods in Ref. \cite{Ozvenchuk:2012kh} have shown that the
solutions are rather close. This holds especially for the case of
the scattering of massive partons where the transport cross section
is not very different from the total cross section as also pointed
out in Ref. \cite{Plumari:2012ep}. Furthermore, we mention that the
definition of the shear viscosity $\eta$ is strictly valid only in
the on-shell limit, however, can be employed also in the DQPM$^*$
since the relaxation times ${\bar \tau}_i$ do not depend on the
masses.

%-------------------------------------------------------------------------

\begin{figure}[h!] %tbh!
\begin{center}
\begin{minipage}{14.5pc}
\includegraphics[width=14pc, height=14pc]{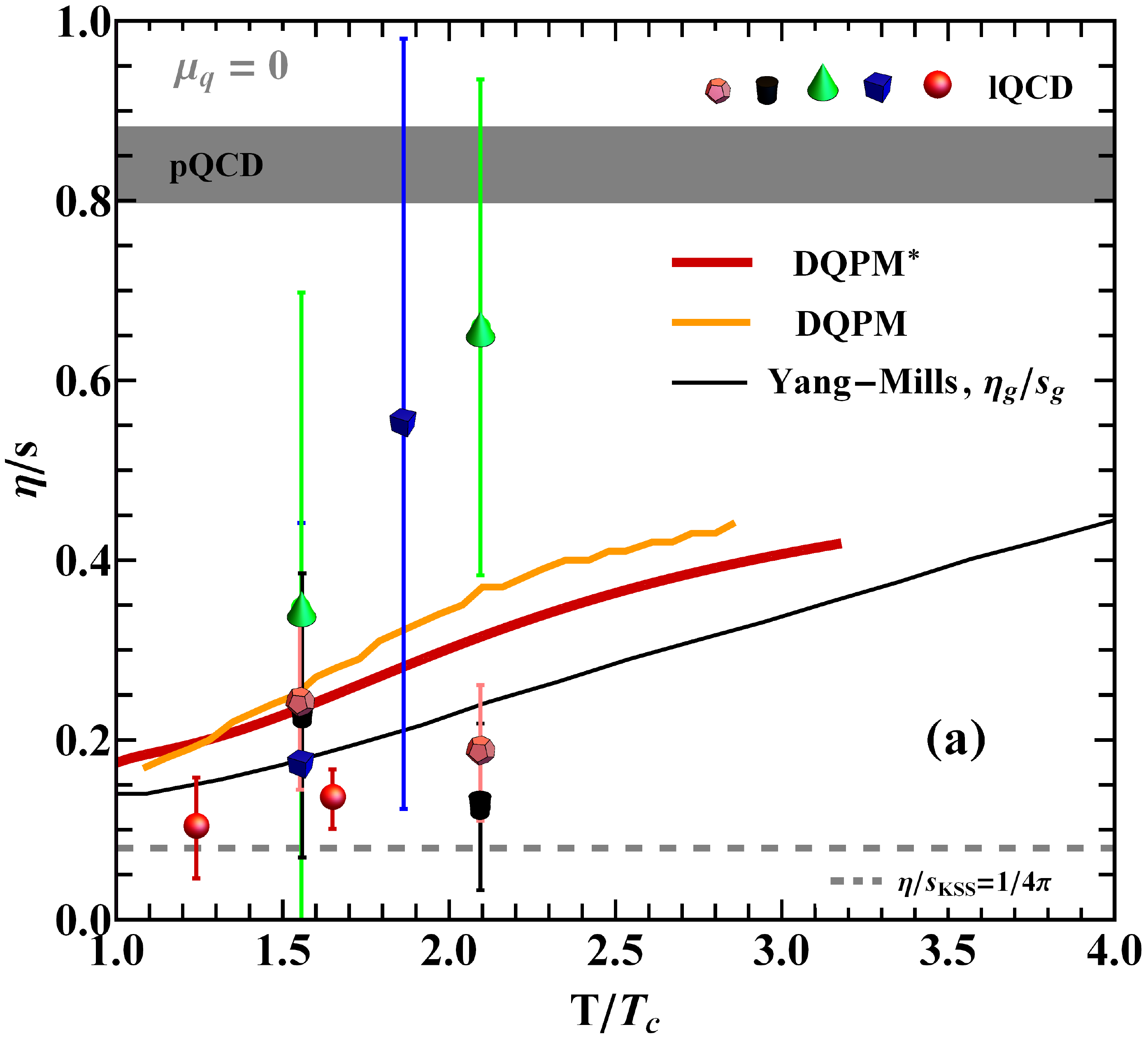}
\end{minipage}
\begin{minipage}{14.5pc}
\includegraphics[width=14pc, height=14pc]{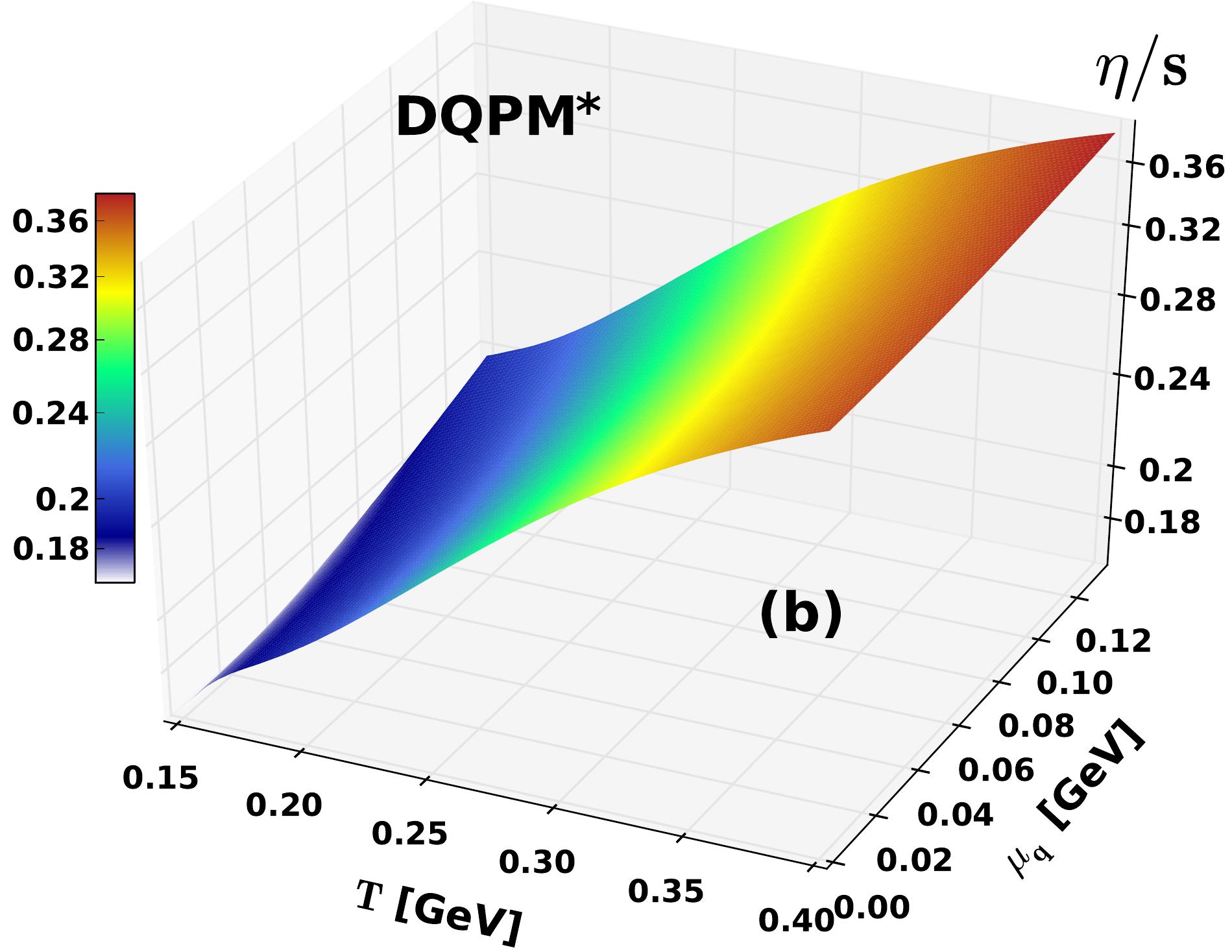}
\end{minipage}
\end{center}
\caption{\emph{(Color online) The shear viscosity to entropy density
ratio $\eta/s$ from different models as a function of temperature
$T$ for $\mu_q =0$ (a) and $\eta/s$ given by the DQPM$^*$ approach
as a function of $(T, \mu_q)$ (b). The orange solid line in (a)
results from the standard DQPM where the parton masses and widths
are independent of momenta \cite{Berrehrah:2014ysa}. The thick red
solid line shows the DQPM$^*$ result using Eqs.(\ref{equ:Sec5.1})
and (\ref{equ:Sec5.2}), where the parton masses and width are
temperature, chemical potential and momentum dependent. The lattice
QCD data for pure $SU(3)$ gauge theory are taken from Ref.
\cite{Meyer:2007ic} (red spheres), from Ref. \cite{Nakamura:2004sy}
(green pyramid and blue cubic), and from Ref. \cite{Sakai:2007cm}
(black cylinder and pink penthagone). The orange dashed line gives
the Kovtun-Son-Starinets lower bound
\cite{Policastro:2001yc,Kovtun:2004de} $(\eta/s)_{KSS} = 1/(4\pi)$.
Finally, the black solid line refers to the calculation of $\eta/s$
in Yang-Mills theory from Ref. \cite{Christiansen:2014ypa}. The figures are taken from Ref. \cite{Berrehrah:2015vhe}.}}
\label{fig:EtaTmu}
\end{figure}

We show the DQPM$^*$ results for $\eta/s$, where $s$ is the DQPM$^*$
entropy density, in Fig.\ref{fig:EtaTmu} (a) as a function of the
temperature. The (upper) orange solid line represents the case of
the standard DQPM where the parton masses and widths are independent
of momenta as calculated in Ref. \cite{Berrehrah:2014ysa}. The thick
red solid line displays the result  using Eqs. (\ref{equ:Sec5.1}) and
(\ref{equ:Sec5.2}), where the parton masses and width are
temperature, chemical potential and momentum dependent. Finally, the
black solid line refers to the calculation of $\eta/s$ in Yang-Mills
theory from the Kubo formula using an exact diagrammatic
representation in terms of full propagators and vertices from Ref.
\cite{Christiansen:2014ypa}.

Fig. \ref{fig:EtaTmu} (a) shows that $\eta/s$ from DQPM$^*$ is in
the range of the lQCD data and significantly lower than the pQCD
limit. As a function of temperature $\eta/s$ shows a minimum around
$T_c$, similar to atomic and molecular systems \cite{Csernai:2006zz}
and then increases slowly for higher temperatures. This behavior is
very much the same as in the standard DQPM (upper orange line) as
shown in Ref. \cite{Ozvenchuk:2012kh}. Therefore, the produced QGP
shows features of a strongly interacting fluid unlike a weakly
interacting parton gas as had been expected from perturbative QCD
(pQCD). The minimum of $\eta/s$ at $T_c = 158$ MeV is close to the
lower bound of a perfect fluid with $\eta/s = 1/(4 \pi)$
\cite{Policastro:2001yc,Kovtun:2004de} for infinitely coupled
supersymmetric Yang-Mills gauge theory (based on the AdS/CFT duality
conjecture). This suggests the ''hot QCD matter'' to be the ''most
perfect fluid'' \cite{Csernai:2006zz}. Furthermore, the ratio
$\eta/s$ in DQPM$^*$ is slightly larger than in the pure gluonic
system (solid black line) due to a lower interaction rate of quarks
and antiquarks relative to gluons.

The explicit dependencies of $\eta/s$ on $T$ and $\mu_q$ are shown
in Fig. \ref{fig:EtaTmu} (b) where $\eta/s$ is seen to increase
smoothly for finite but small $\mu_q$. We point out again that
extrapolations to larger $\mu_q$ become increasingly uncertain.

The bulk viscosity (defined in Ref. \cite{Chakraborty:2010fr} for
the on-shell case) reads in the relaxation time approximation (RTA)
for the case of off-shell DQPM$^*$ partons as:
{\setlength\arraycolsep{0pt}
\begin{eqnarray}
\label{zeta}
\zeta (T,\mu_q) = & & \ \frac{1}{9 T} d_g  \!\!\int\!\! \frac{d^3 p}{(2\pi)^3} \ \!\!\int\!\!
\frac{d\omega}{2 \pi} \omega \ \bar \tau_g (T, \mu_q) \ f_g(\omega/T) \ \rho_{g}
(\omega, {\boldsymbol p}) \ \Theta(P^2)   \frac{1}{\omega^2} \ F_g(\omega, {\boldsymbol p})
\nonumber\\
& & {} + \frac{1}{9 T} \frac{d_q}{6} \!\!\int\!\! \frac{d^3
p}{(2\pi)^3} \ \!\!\int\!\! \frac{d \omega}{2 \pi} \omega \  \Bigg[
\sum_q^{u,d,s} \bar \tau_q (T, \mu_q) f_q((\omega-\mu_q)/T)
\rho_{q}(\omega,{\boldsymbol p})
\nonumber\\
& & {} + \sum_{\bar q}^{\bar u,\bar d,\bar s}
\bar \tau_{\bar q} (T, \mu_q) f_{\bar q}((\omega+\mu_q)/T) \ \rho_{\bar
q}(\omega,{\boldsymbol p}) \  \Bigg] \ \Theta(P^2)  \frac{1}{\omega^2}  \ F_q(\omega, {\boldsymbol p}) ,
\end{eqnarray}} with
\begin{equation} \label{extra} F_i(\omega, {\boldsymbol p})
\Bigg[{\boldsymbol p}^2 - 3 c_s^2 \Bigg(\omega^2 - T^2 \frac{d
M_i^2}{d T^2} \Bigg) \Bigg]^2
\end{equation}
and essentially depends on   the mass derivatives $\partial M_i^2 /
\partial T^2$,  the temperature $T$, and the speed of
sound squared $c_s^2$. All these quantities are accessable within
the DQPM$^*$ such that the results for the bulk viscosity again do
not imply any new parameter.

\begin{figure}[h!] %tbh!
\begin{center}
\begin{minipage}{14.5pc}
\includegraphics[width=14pc, height=14pc]{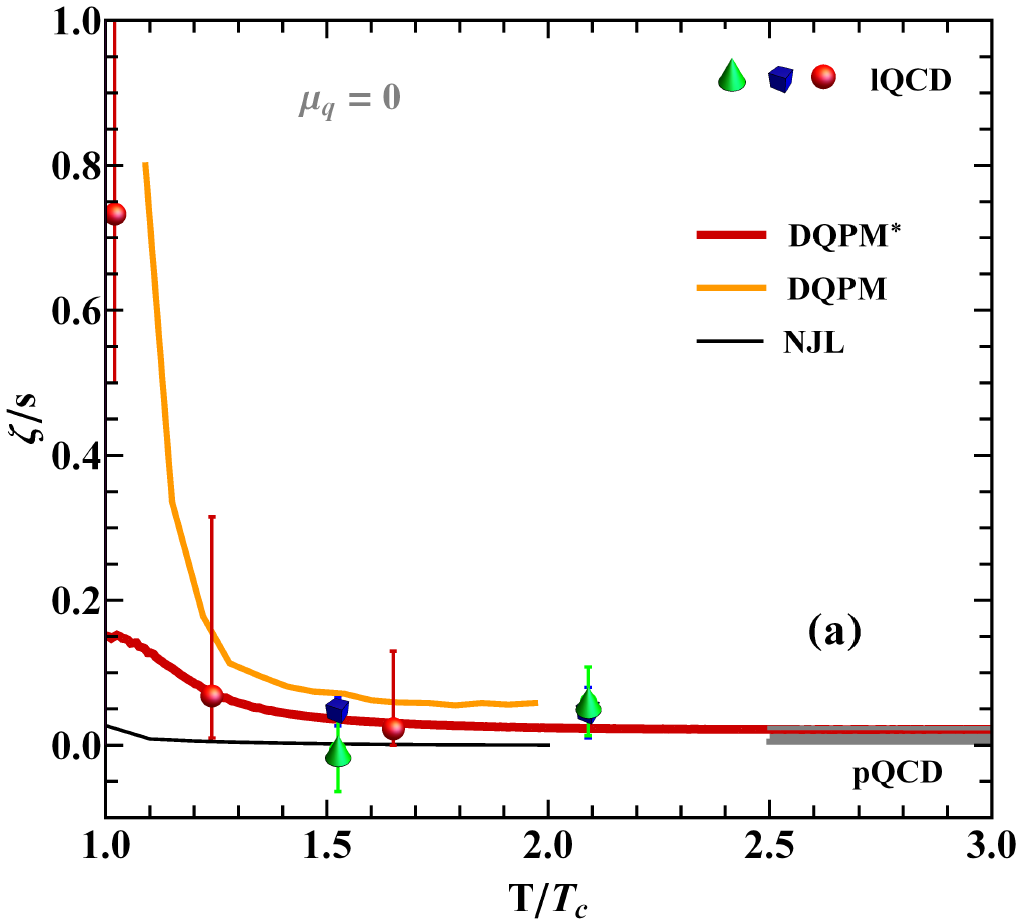}
\end{minipage}
\begin{minipage}{14.5pc}
\includegraphics[width=14pc, height=14pc]{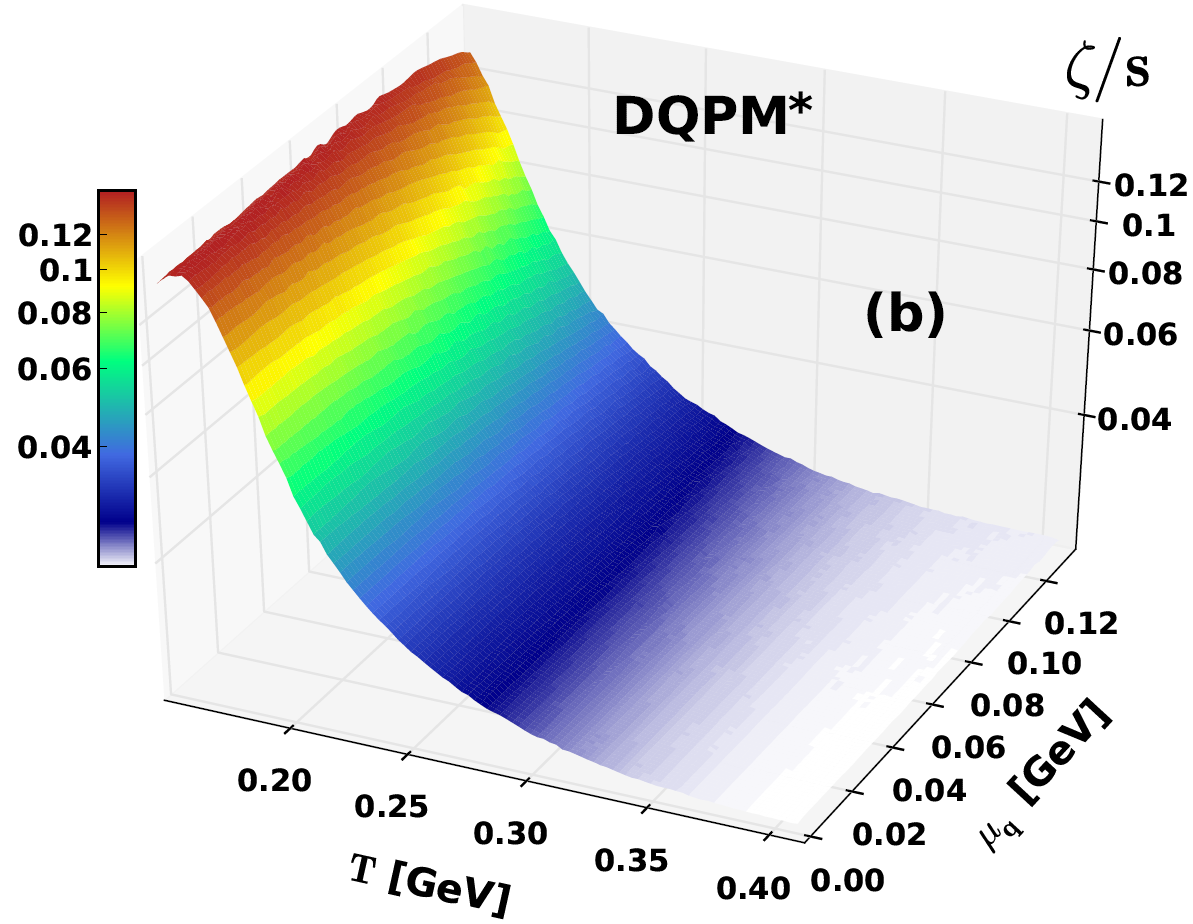}
\end{minipage}
\end{center}
\caption{\emph{(Color online) The bulk viscosity to entropy density
ratio $\zeta/s$ from DQPM$^*$  as a function of temperature $T$ for
$\mu_q =0$ (a) and $\zeta/s$ given by the DQPM$^*$ approach as a
function of $T$ and $\mu_q$ (b). The orange solid line in (a)
results from the standard DQPM where the parton masses and widths
are independent of momenta \cite{Berrehrah:2014ysa}. The lattice QCD data
points for pure $SU(3)$ gauge theory are taken from Ref. \cite{Meyer:2007ic} (red spheres),
\cite{Sakai:2007cm} (blue cubic) and from Ref. \cite{Nakamura:2004sy} (green pyramid).
Finally, the black solid line in (a) refers to the calculation of $\zeta/s$ from the
Nambu-Jona-Lasinio model for $SU(3)_f$ from Ref. \cite{Marty:2013ita}.}} \label{fig6}
\end{figure}

The bulk viscosity (divided by the entropy density $s$) from the
DQPM$^*$  is displayed in Fig. \ref{fig6} (a) and shows a very
different temperature dependence  than $\eta/s$. Indeed, for high
temperatures we find the limit $\zeta/s \to 0$. Moreover, the
behavior around $T_c$ shows a peak in lQCD as well as in the DQPM
and  DQPM$^*$ which is essentially due to the  derivative $\partial
M_i^2 / \partial T^2$ in Eq. (\ref{zeta}). Accordingly, the infrared
enhancement in the DQPM$^*$ masses is mandatory to achieve a maximum
in the bulk viscosity $\zeta(T,\mu_q)$ to entropy ratio $\zeta/s$
close to $T_c$ in line with lQCD. This enhancement close to $T_c$ is
lower in the DQPM$^*$ as in the DQPM probably due to a lower
infrared enhancement in the coupling squared. Note, however, that
such an enhancement does not show up in the NJL calculations for
$\zeta/s$ from Ref. \cite{Marty:2013ita} (black solid line in (a)).
The explicit dependencies of $\zeta/s$ on $T$ and $\mu_q$ from the
DQPM$^*$ are shown in Fig. 6 (b).

%--------------------------------------------------------------------------------------------------------------------------------------------------------------------------------
\subsection{Electric conductivity}
\label{etaOvs2}
%--------------------------------------------------------------------------------------------------------------------------------------------------------------------------------

Whereas the shear and bulk viscosities depend on the properties of
quarks, antiquarks and gluons the electric conductivity $\sigma_e$
only depends on electrically charged quarks and antiquarks and thus
provides independent information on the response of the QGP to
external electric fields \cite{Cassing:2013iz,Steinert:2013fza}. It
probes exclusively the fermion properties (as in case of the quark
susceptibilities) and the interaction strength with gluons enters
only indirectly via the total width of the quarks and antiquarks.
The electric conductivity $\sigma_e$ is also important for the
creation of electromagnetic fields in ultra-relativistic
nucleus-nucleus collisions from partonic degrees-of-freedom, since
$\sigma_e$ specifies the imaginary part of the electromagnetic
(retarded) propagator and leads to an exponential decay of the
propagator in time $\sim \! \exp(-\sigma_e (t-t'))$. Furthermore,
$\sigma_e$ also controls the photon spectrum in the long wavelength
limit \cite{Linnyk:2015tha}.

We recall that the dimensionless ratio $\sigma_e/T$ in the
quasiparticle approach is given by the relativistic Drude formula
\cite{Cassing:2013iz,Marty:2013ita,Steinert:2013fza},
{\setlength\arraycolsep{-1pt}
\begin{equation}
\label{equ:Sec6.1}  \sigma_e (T,\mu_q) = \sum_{f,\bar{f}}^{u,d,s}
\frac{e_f^2 \ n_f^{\textrm{off}} (T, \mu_q)}{\bar \omega_f (T,
\mu_q) \ \bar \gamma_f (T, \mu_q)},
\end{equation} with \begin{equation}
  \bar \omega_f (T, \mu_q)=
\left(n_{f}^{\textrm{off}} (T,\mu_q)\right)^{-1}   \displaystyle
\int \frac{d^3p}{(2 \pi)^3} \ \frac{d \omega}{(2 \pi)} \ \omega^2 \
\rho_f (\omega, {\boldsymbol p}) \ f_f((\omega\pm \mu_q)/T) \ ,
\end{equation}}
where the quantity $\bar \omega_q(T,\mu_q)$ is the quark (antiquark)
energy averaged over the equilibrium distributions at finite $T$ and
$\mu_q$ while $\bar \gamma_q(T,\mu_q)$ is the averaged quark width,
as given in Eq. (\ref{equ:Sec5.3}).

\begin{figure}[h!] %tbh!
\begin{center}
\begin{minipage}{14.5pc}
\includegraphics[width=14.5pc, height=14pc]{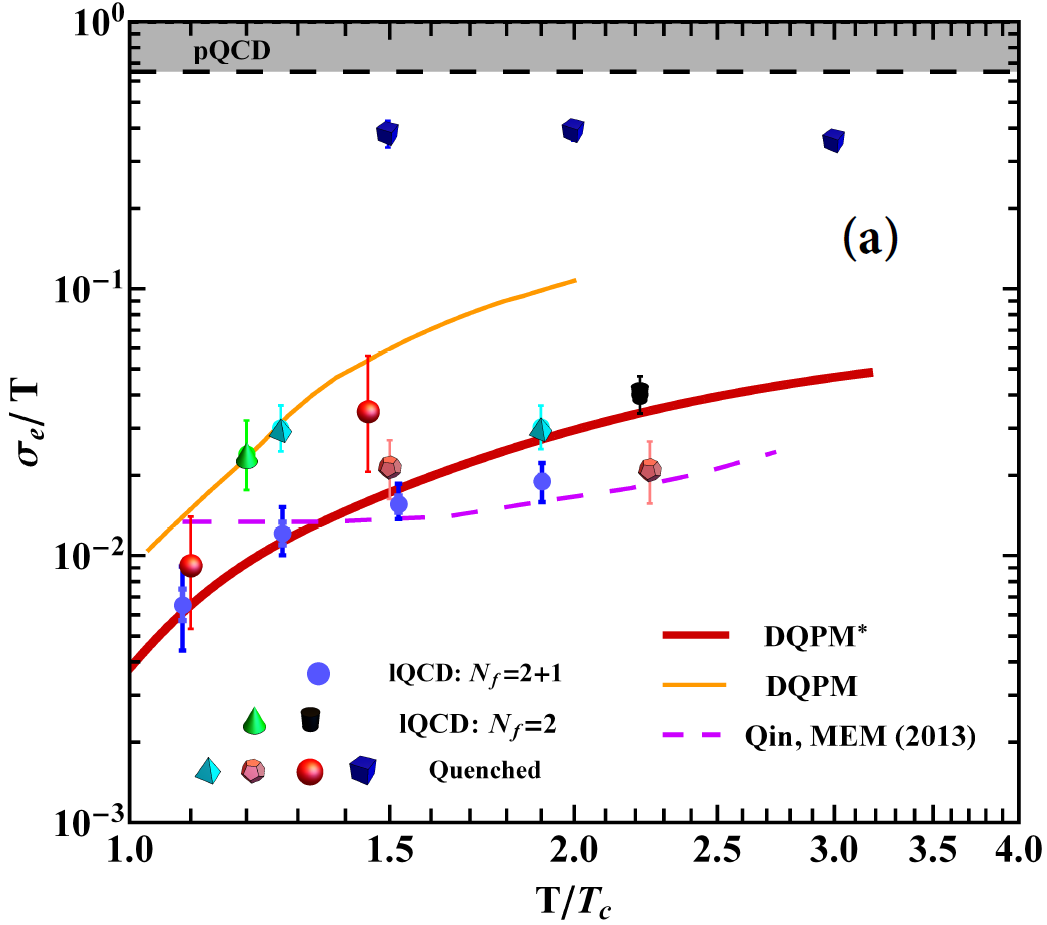}
\end{minipage}
\begin{minipage}{14.5pc}
\includegraphics[width=14.5pc, height=14pc]{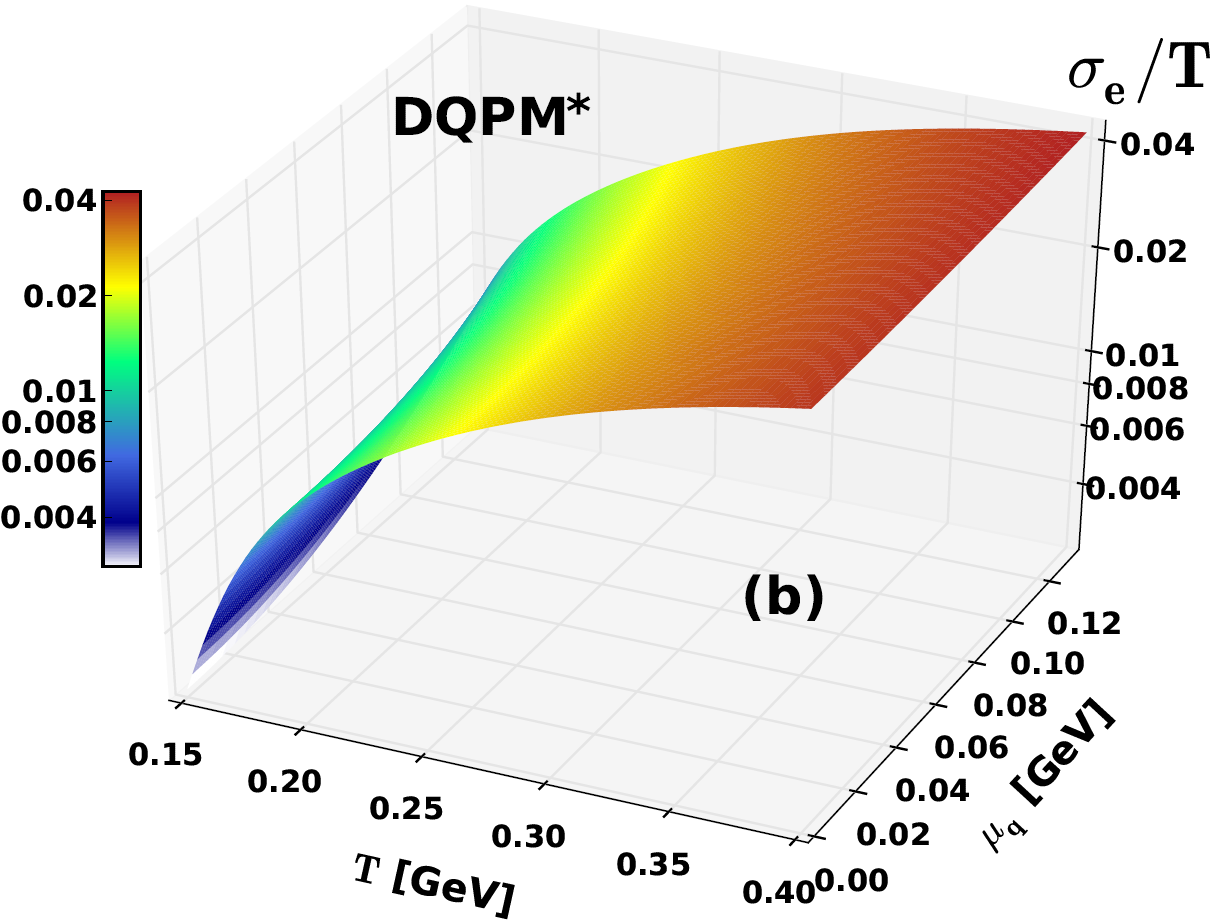}
\end{minipage}
\end{center}
\caption{\emph{(Color online) The ratio $\sigma_e/T$ from different
models as a function of temperature $T$ for $\mu_q =0$ (a) and
$\sigma_e/T$ given by the DQPM$^{\star}$ approach as a function of
$(T, \mu_q)$ (b). The orange thin solid line in (a) results from the
standard DQPM where the parton masses and widths are independent of
momenta \cite{Berrehrah:2014ysa}. The red thick solid line shows the
DQPM$^*$ result  using Eqs.(\ref{equ:Sec6.1}), where the parton
masses and width are temperature, chemical potential and momentum
dependent. The lattice QCD data are taken from Ref.
\cite{Ding:2010ga} (red spheres), Ref. \cite{Aarts:2007wj} (pink
pentagon), Ref. \cite{Gupta:2003zh} (blue cubic), Ref.
\cite{Kaczmarek:2013dya} (Cyan pyramid), Ref. \cite{Brandt:2013fg}
(green cone), Ref. \cite{Buividovich:2010tn} (black cylinder), Ref.
\cite{Aarts:2014nba} (blue disk). Qin, MEM (2013) refers to Ref.
\cite{Qin:2013aaa} where a Dyson-Schwinger approach is used. The
electric charge is explicitly multiplied out using $e^2 \approx 4
\pi/137$. The average charge squared is $C_{EM} = 8 \pi \alpha/3$
with $\alpha \approx 1/137$. Note that the pQCD result at leading
order beyond the leading log \cite{Arnold:2003zc} is $\sigma_e/T
\approx 5.97/e^2 \approx 65$. The figures are taken from
Ref. \cite{Berrehrah:2015vhe}.}} \label{fig:SigeTmu}
\end{figure}

The actual results for $\sigma_e/T$ are displayed in Fig.
\ref{fig:SigeTmu} (a) in terms of the thick red solid line in
comparison to recent lQCD data from Refs.
\cite{Ding:2010ga,Aarts:2007wj,Gupta:2003zh,Kaczmarek:2013dya,Brandt:2013fg,Buividovich:2010tn,Aarts:2014nba}
and the result from previous studies within the DQPM
\cite{Marty:2013ita} (thin orange line). Again we find a minimum in
the partonic phase close to $T_c$ and a rise with the temperature
$T$. The explicit dependencies of $\sigma_e/T$ on $T$ and $\mu_q$,
shown in Fig. \ref{fig:SigeTmu} (b), is also increasing smoothly for
finite but small $\mu_q$. We finally note that the lower values for
$\sigma_e/T$ in the DQPM$^*$ relative to the DQPM result from using
the relativistic Drude formula (\ref{equ:Sec6.1}) instead of its
nonrelativistic counterpart.

\section{Differential partonic cross sections and transport
coefficients for charm quarks in the hot medium}
 Since the properties of charm degrees of freedom $Q$ cannot be determined from the usual thermodynamic quantities
due to their large mass (and low occupation probability in thermal equilibrium) we calculate the
charm interaction rates in the partonic medium dynamically via explicit cross sections.
We will use the standard DQPM with momentum-independent selfenergies to evaluate
the charm scattering with the 'dressed' ($u,d,s$) partons.

%---------------------------------------------------------------------------------------------------------------------------------------------
\subsection{$q Q$ and $g Q$ \ elastic scattering at finite $T$ and $\mu_q$}
\label{qgQProcess}
%---------------------------------------------------------------------------------------------------------------------------------------------
\vskip -2mm

The process $q Q \rightarrow q Q$ is calculated here to lowest order
in the perturbation expansion using the extended Feynman rules for
massless quarks in Politzer's review \cite{DavidPolitzer1974129} for
the case of finite masses and widths. The color sums are evaluated
using the techniques discussed in Ref. \cite{DavidPolitzer1974129} ;
the spin sums will be discussed below. Contrary to the case of
massless gluons where the \emph{``Transverse gauge''} is used, the
\emph{``Lorentz covariance''} is used for the case of massive gluons
here since a finite mass in the gluon propagator allows to fix the
0'th components of the gluon fields $A^0_a$ ($a=1, \cdots, 8$) by
the spatial degrees of freedom  $A^k_a (k=1,2,3)$. Furthermore, the
divergence encountered in the $t$-channel (Refs.
\cite{PhysRevD.17.196,Combridge1979429,Combridge1977234}) -- when
calculating the total cross sections $\sigma^{qQ}$ and $\sigma^{gQ}$
-- is cured self-consistently since the infrared regulator is given
by the finite DQPM gluon mass (and width). In our calculations, we
have developed two different models, the so-called DpQCD (Dressed
pQCD) and IEHTL (Infrared Enhanced HTL) model. In the first we
consider only massive gluons and light and heavy quarks with masses
given by the DQPM pole masses, whereas both the DQPM masses and
widths are considered in the IEHTL model \cite{Berrehrah:2015ywa}.
In the following we will only report on results obtained within the
DpQCD since a finite width in the charm and light quark spectral
functions was shown to lead to very moderate modifications of the
results \cite{Berrehrah:2015ywa}.

\begin{figure}[h!]
\begin{center}
      \includegraphics[height=3cm, width=3.5cm]{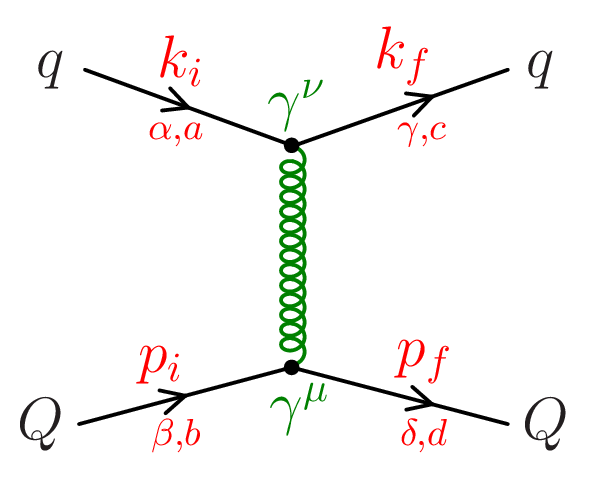}
%      \hspace{0.8cm}
      \includegraphics[height=3cm, width=9cm]{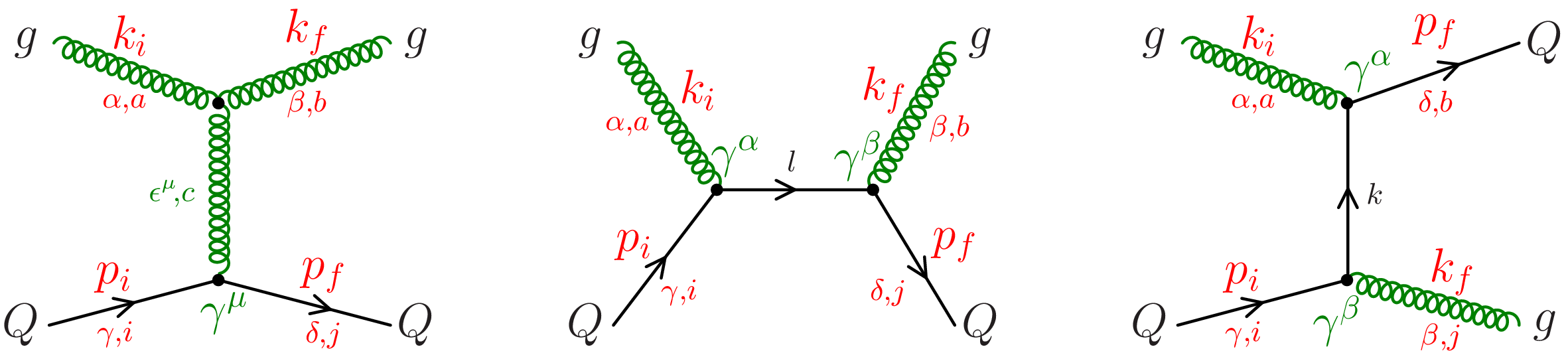}
\end{center}
\caption{\emph{(Color online) Feynman diagrams for the $q Q
\rightarrow q Q$ and $g Q \rightarrow g Q$ scattering processes. Latin
(Greek) subscripts denote colour (spin) indices. $k_i$, resp. $p_i$
($k_f$, resp. $p_f$) denote the initial (final) 4-momentum of the
light quark or the gluon, resp. the heavy quark $Q$. The invariant energy
squared is given by $s=(p_i + k_i)^2$, $t=(p_i - p_f)^2$, and $u=(p_i -
k_f)^2$. The figure is taken from Ref. \cite{Berrehrah:2015ywa}. } \label{fig:qgQFeynmanDiagram}}
\end{figure}

The elementary Feynman
diagrams for the $qQ$ and $gQ$ elastic scattering at order
$O(\alpha_s)$ are illustrated in Fig.  \ref{fig:qgQFeynmanDiagram}.

%\begin{widetext}
%\end{widetext}

For on-shell $qQ$ elastic
scattering, the $t$-channel invariant squared amplitude - averaged
over the initial spin and color degrees of freedom and summed over
the final state spin and color - $\mathcal{M}_t$ is given by
%\begin{widetext}
{\setlength\arraycolsep{0pt}
\begin{eqnarray}
\label{equ:Sec3.2} \hspace*{-0.5cm} \sum |\mathcal{M}_t|^2 \!\! =
\!\! \frac{4 g^4}{9 (t - m_g^2)^2} \biggl[ (s-M_Q^2-m_q^2)^2 + (u -
M_Q^2 - m_q^2)^2 + 2 (M_Q^2 + m_q^2)t \biggr],
\end{eqnarray}}
%\end{widetext}
where $m_q$ ($M_Q$) is the light quark (heavy quark) mass and $m_g$
is the DQPM exchanged gluon mass.

In the off-shell picture we take into account not only the finite
masses of the partons, but also their spectral functions, i.e. their
finite widths. Since the light quark and heavy quark masses change
before and after the scattering ('quasi-elastic' process) we
introduce the mass $m_q^i$ for the initial $q$ and $m_q^f$ for the
final $q$, and allow for different masses of the heavy quark,
$M_Q^i$ for the initial $Q$ and $M_Q^f$ for the final $Q$. The
squared amplitude -- averaged over the initial spin and color
degrees of freedom and summed over the final state spin and color --
gives:
%\begin{widetext}
{\setlength\arraycolsep{0pt}
\begin{eqnarray}
\label{equ:Sec3.3}
\sum |\mathcal{M}|^2 = & & \frac{2 g^4}{9 \bigl[(t-m_g^2)^2+4\gamma_g^2 q_0^2\bigr]} \times
\nonumber\\
& & {} \biggl[ 4 \left(p_f^{\mu} p_i^{\nu} + p_i^{\mu} p_f^{\nu} + g^{\mu \nu} \frac{t}{2} \right)
\biggr] \biggl[ 4 \left(k_{f,\mu} k_{i,\nu} + k_{i,\mu} k_{f,\nu} + g_{\mu \nu} \frac{t}{2} \right) \biggr],
\end{eqnarray}}
%\end{widetext}
where we have incorporated the DQPM propagators (i.e. $t_{\pm}^{*} =
t - m_g^2 \pm 2 i \gamma_g q_0$, where $m_g$, $\gamma_g$ are,
respectively, the effective gluon mass and total width at
temperature $T$ and quark chemical potential $\mu_q$ while $q^0 =
p_f^0 - p_i^0 = k_f^0 - k_i^0$ is the gluon energy in the
$t$-channel). Thus the divergence in the gluon propagator in the $t$-channel is regularized.

The relative contribution of the off-shell partons to the pQCD cross
section is expected to change due to different kinematical
thresholds and to the changes in the matrix element- corresponding
to the diagram in Fig. \ref{fig:qgQFeynmanDiagram}. The off-shell
kinematical limits for the momentum-transfer squared $t$ and the
expressions of the Mandelstam variables in the case of off-shell
heavy quark scattering are given in Ref. \cite{Berrehrah:2013mua}.
Inspite of these expectations the actual results in Ref. \cite{Berrehrah:2015ywa}
show that the finite width of the partons has only a minor impact
on the charm scattering with the bulk partons.

\begin{figure}[h!]
       \begin{center}
       \includegraphics[width=6.2cm, height=7cm]{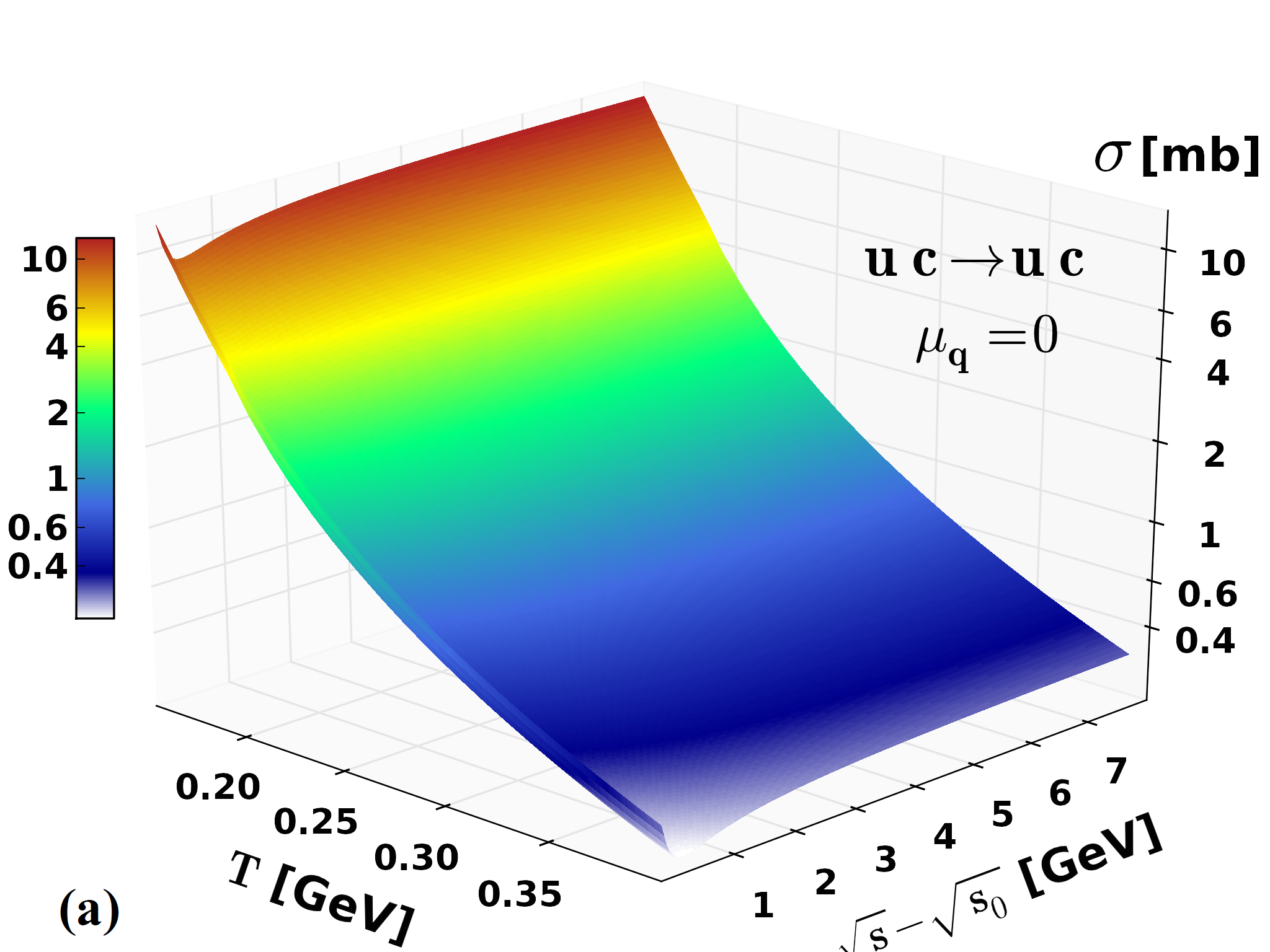}
       \includegraphics[width=6.2cm, height=7cm]{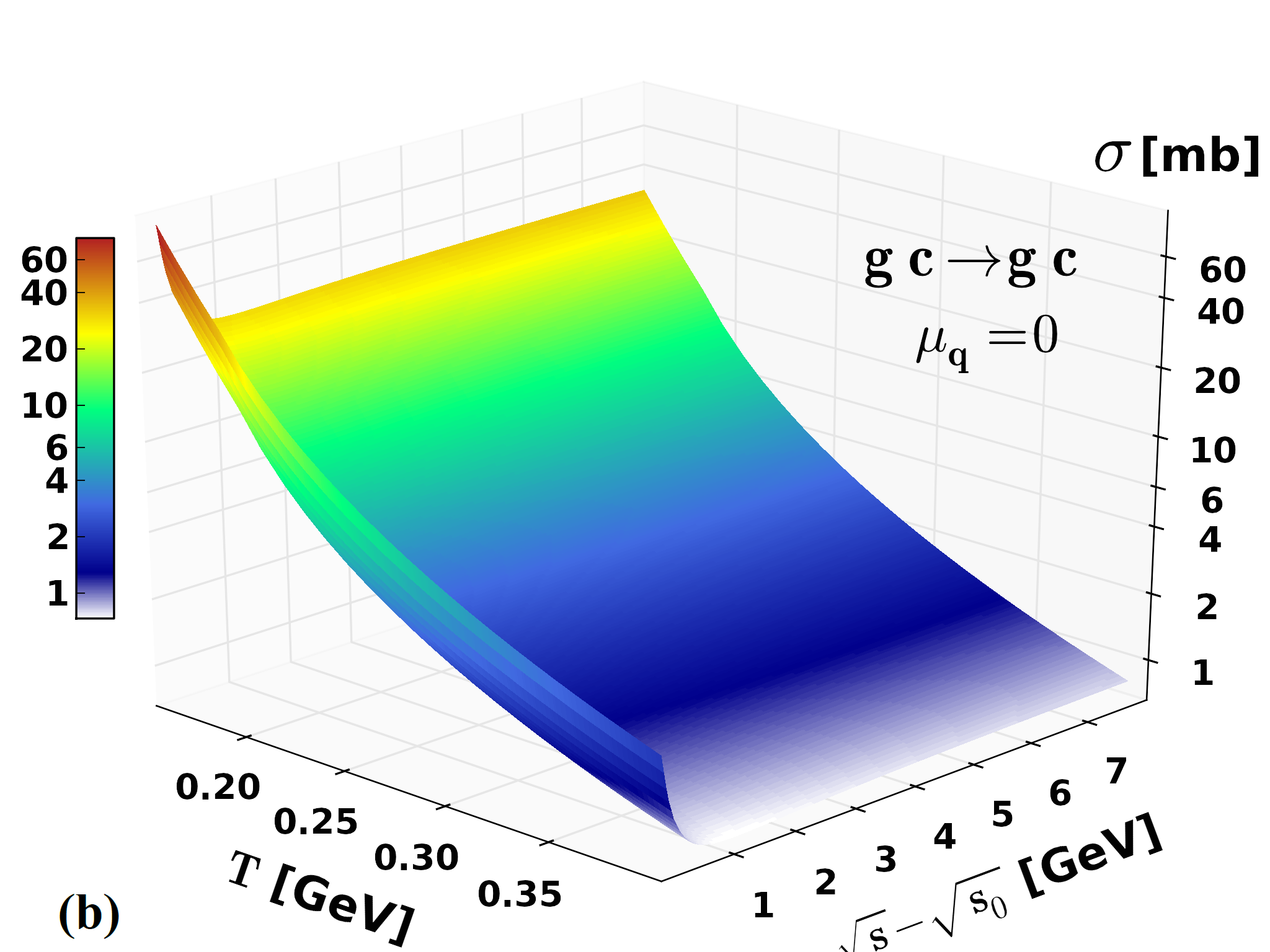}
       \end{center}
\vspace*{-0.2cm}
\caption{\emph{(Color online) Elastic cross section of $u c \rightarrow u c$ (a) and
$g c \rightarrow g c$ (b) scattering as a function of the temperature $T$ and the invariant
energy above threshold $\sqrt{s} - \sqrt{s_0} $, where $\sqrt{s_0}$ is the threshold energy,
for on-shell partons as described by the DpQCD approach at $\mu_q = 0$.
The figures are taken from Ref. \cite{Berrehrah:2015ywa}.
} \label{fig:Sigugc-DpQCD-IEHTL_vsTS}}
\end{figure}

Figs. \ref{fig:Sigugc-DpQCD-IEHTL_vsTS} (a) and (b) show explicitly
the temperature and $\sqrt{s}$ dependences of the $uc$ and $gc$
elastic cross sections at $\mu_q = 0$, as described in the DpQCD
approach. We find that an increasing medium temperature $T$ leads
to an increase of the thermal gluon mass (infrared regulator) and
hence to a decrease of the DpQCD $uc$ and $gc$ elastic cross
sections. We recall that the effective gluon mass is roughly proportional to $T$
for temperatures above $0.2$ GeV. The large enhancement of the total
cross section for temperatures close to $T_c (\mu_q)$, furthermore, can be traced
back to the infrared enhanced coupling.

%----------------------------------------------------------------------------------------------------------------------------------------------------------------------------
\subsection{Heavy quark interaction rates in a medium at finite $T$ and $\mu_q$}
\label{HQRates}
%----------------------------------------------------------------------------------------------------------------------------------------------------------------------------
\vskip -2mm

Using the elastic cross section for $q (\bar{q})Q$ and $g Q$
collisions, for on- and off-shell partons -- as calculated in
Sec. \ref{qgQProcess} --  we evaluate the interaction rate  of a
heavy quark $Q$ with momentum $\boldsymbol{p}$ and energy $E$ propagating
through a QGP in thermal and chemical equilibrium at a given
temperature $T$ and quark chemical potential $\mu_q$. The occupation numbers of the light
quarks/antiquarks of the plasma are described by a Fermi-Dirac
distribution $f_{q,\bar{q}} (\boldsymbol{q}) = (e^{(E_q \mp \mu_q)/
T} + 1)^{-1}$ whereas the gluons follow a Bose-Einstein distribution
$f_{g} (\boldsymbol{q}) = (e^{E_g/ T} - 1)^{-1}$.

For on-shell particles (DpQCD) and in the reference system in
which the heavy quark has the velocity $\boldsymbol{\beta} = \boldsymbol{p}/E$ the
(on-shell)  interaction rate $R^{\textrm{on}} (\boldsymbol{p}) = d
N_{coll}^{2 \rightarrow 2}/dt$ for $2 \rightarrow 2$ collisions is
given by \cite{Berrehrah:2014kba},
{\setlength\arraycolsep{0pt}
\begin{eqnarray}
\label{equ:Sec4.1}
R^{\textrm{on}} (\boldsymbol{p}, T, \mu_q) = \displaystyle \sum_{q,\bar{q},g}
\frac{M_Q}{16 (2 \pi)^4 E} \int \frac{q^3 m_0^{\textrm{on}} (s) f_r (\boldsymbol{q})}{s \ E_q} \ d q,
\end{eqnarray}}
where the sum is over the
light quarks/antiquarks and gluons of the medium. In Eq.
(\ref{equ:Sec4.1})  $f_r (\boldsymbol{q})$ is the invariant distribution of
the plasma constituents in the rest frame of the heavy quark, given
for the quark/antiquark by:
{\setlength\arraycolsep{0pt}
\begin{eqnarray}
\label{equ:Sec4.2}
\int d\Omega\ f_r (\boldsymbol{q})= 2\pi \int_{-1}^{1} d cos\theta_r \ \frac{1}{e^{(u^0 E_q - u \ q \cos \theta_r \mp \mu_q)/T} + 1},
\end{eqnarray}}
with $u\equiv (u^0,\boldsymbol{u}) = \frac{1}{M_Q}(E,-\boldsymbol{p})$ being the
fluid 4-velocity measured in the heavy-quark rest frame, while
$\theta_r$ is the angle between $\boldsymbol{q}$ and $\boldsymbol{u}$.
The quantity $m_0^{\textrm{on}} (s)$ in Eq. (\ref{equ:Sec4.1}) is related to the
transition amplitude $|\mathcal{M}_{2,2}|^2$ of the collision
$q(\bar{q},g) Q \rightarrow q(\bar{q},g) Q$ by
{\setlength\arraycolsep{0pt}
\begin{eqnarray}
\label{equ:Sec4.3}
m_0^{\textrm{on}} (s) = \frac{1}{2 p_{cm}^2 (s)} \!\! \int_{- 4 p_{cm}^2}^0 \
\frac{1}{g_Q g_p} \sum_{i,j} \sum_{k,l} |\mathcal{M}_{2,2} (s, t; i, j | k, l)|^2 \ d t,
\end{eqnarray}}
with $p_{cm} = (q \ M_Q)/\sqrt{s}$ denoting the momentum of the
scattering partners in the c.m. frame and $g_Q$ ($g_p$)  the
degeneracy factor of the heavy quark (parton).

\begin{figure}[h!] %tbh!
\begin{center}
     \includegraphics[height=4.5cm, width=4.05cm]{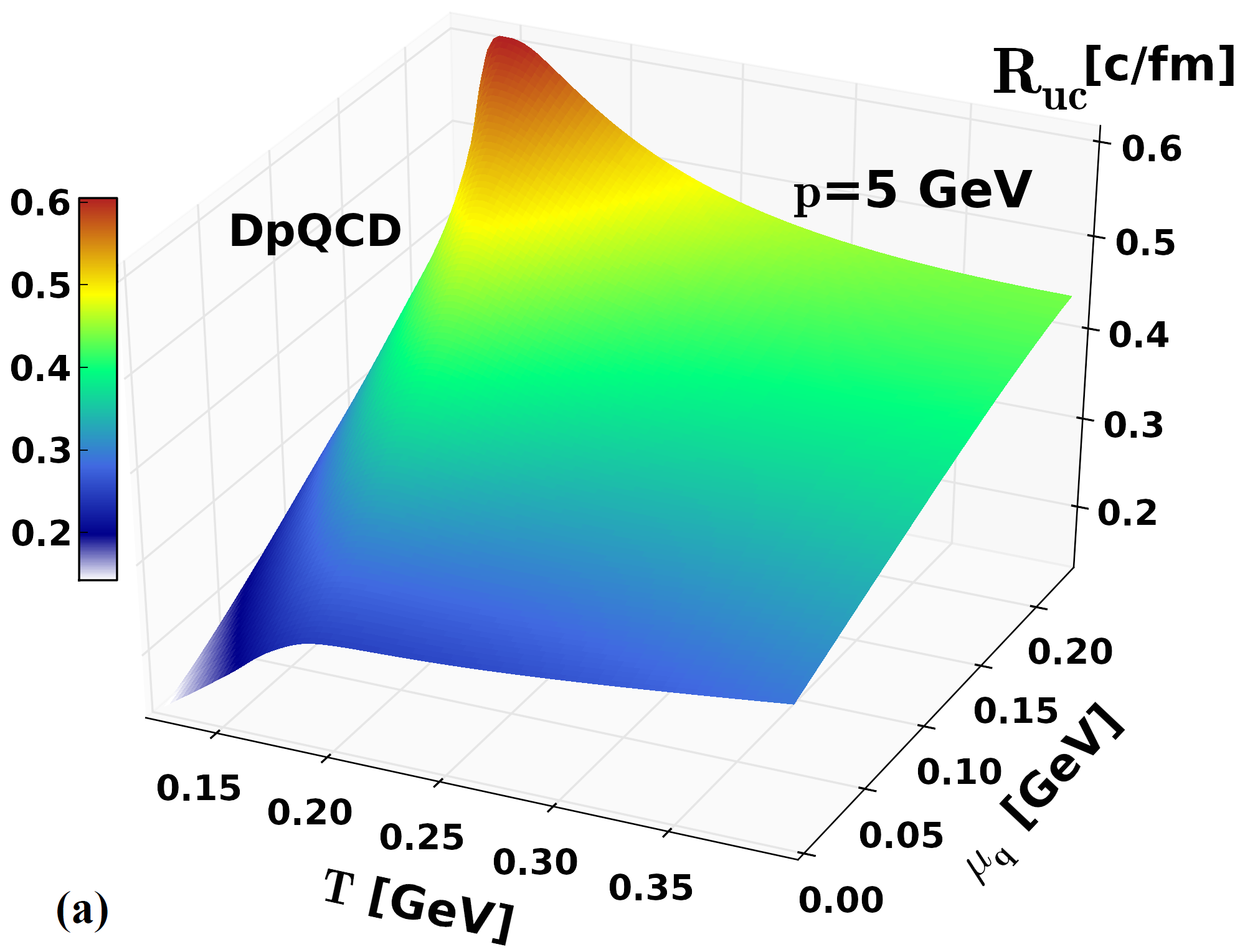}
     \hspace{0.01cm}
     \includegraphics[height=4.5cm, width=4.05cm]{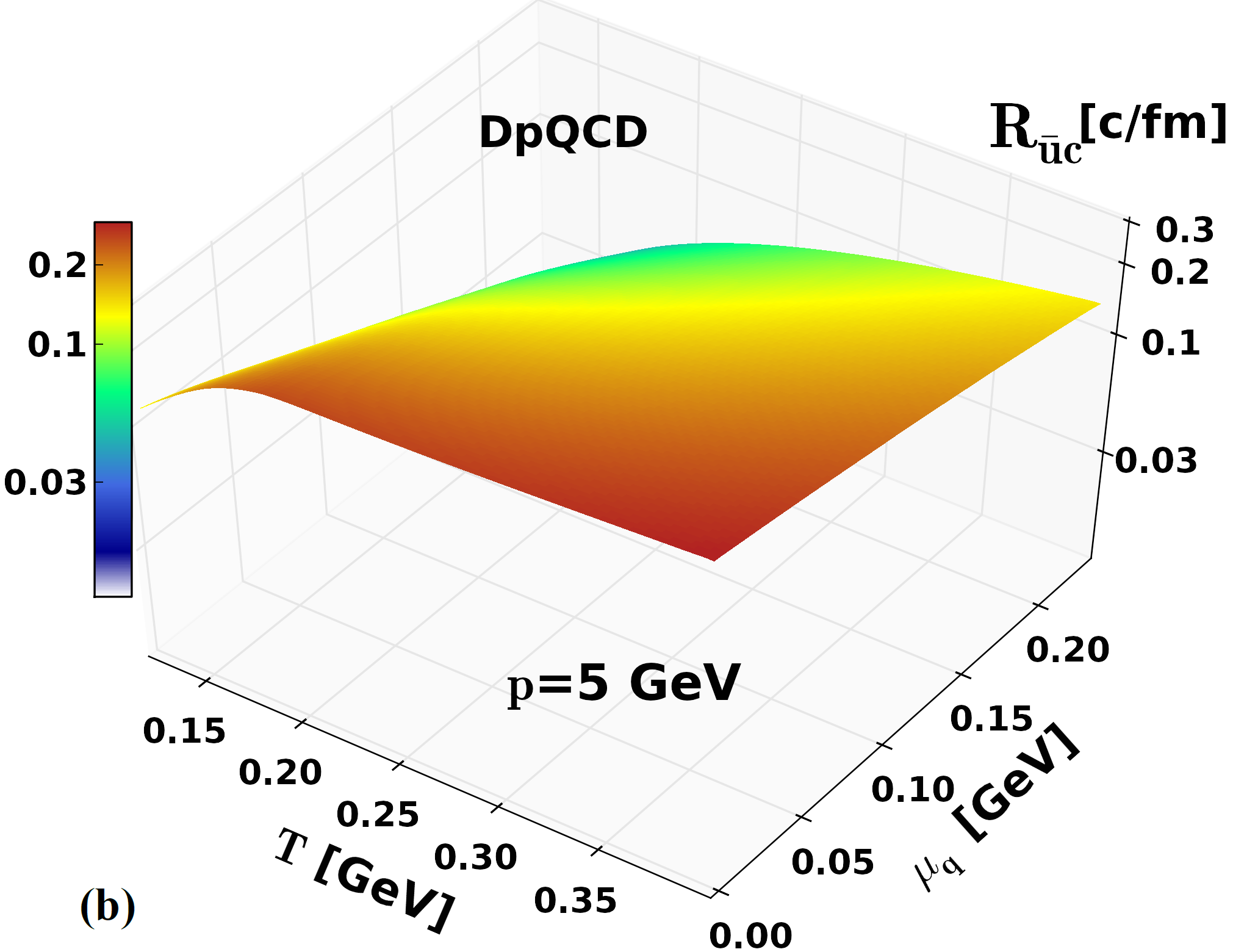}
     \hspace{0.01cm}
     \includegraphics[height=4.5cm, width=4.05cm]{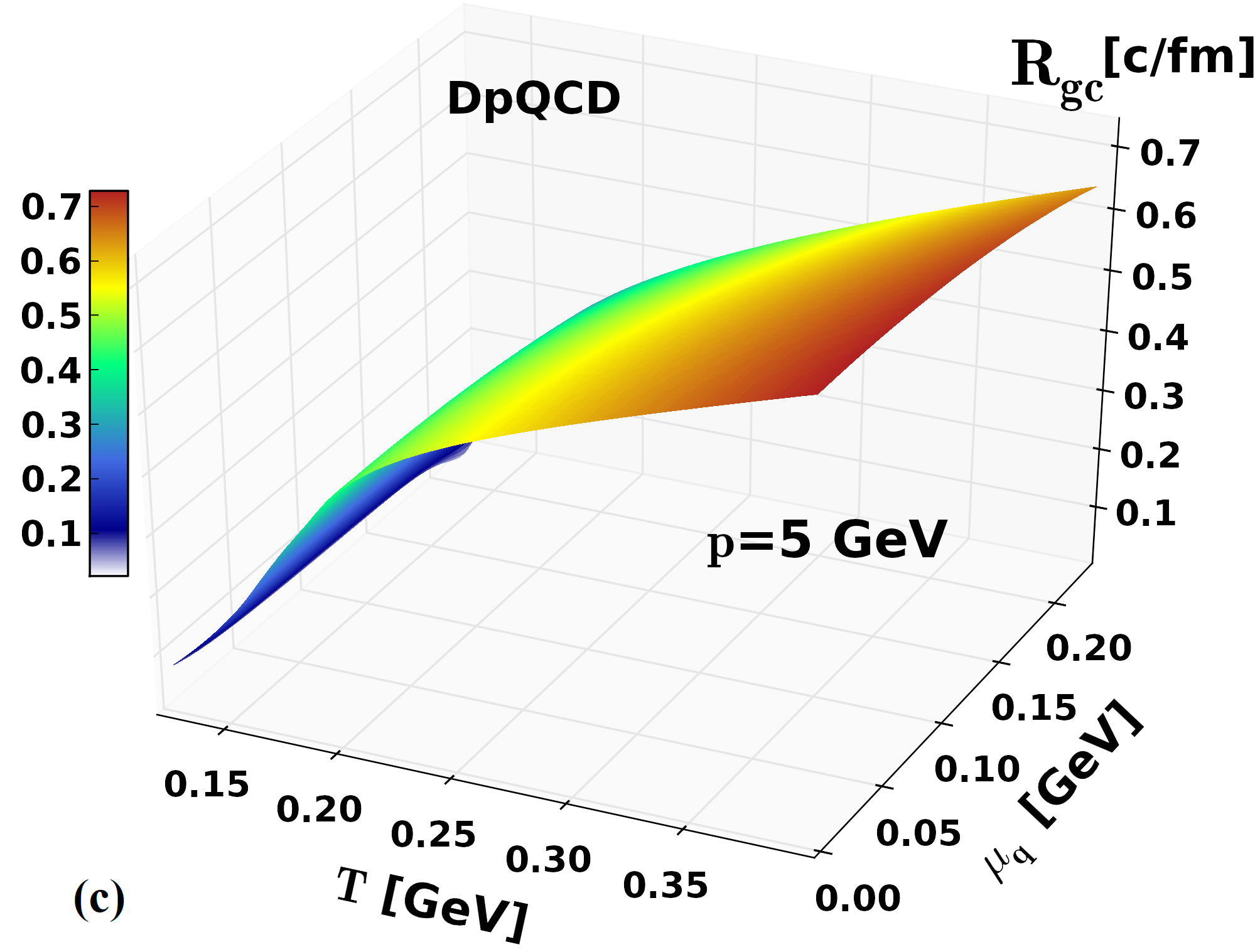}
\vspace{-0.3cm}
\caption{\emph{(Color online) The total elastic interaction rate $R$ of c-quarks in the plasma
rest frame as a function of the temperature $T$ and quark chemical potential $\mu_q$ due to the scattering
with light quarks (a), light antiquarks (b) and gluons (c). The on-shell heavy-quark momentum in all cases is $p = 5$ GeV/c.
The figures are taken from Ref. \cite{Berrehrah:2015ywa}.
}}
\label{fig:cRatevsTmu}
\end{center}
\end{figure}

Due to the different abundances of particle species in a medium at
finite chemical potential, it is interesting to study the variation
of the heavy-quark interaction rates with the quarks/antiquarks and
gluons independently. Figs. \ref{fig:cRatevsTmu}-(a), (b) and (c)
illustrate the dependence of the heavy-quark collisional rates with
quarks, antiquarks and gluons of a medium at finite temperature $T$
and quark chemical potential $\mu_q$  for an intermediate
heavy-quark momentum ($p = 5$ GeV/c).

For the case of gluons and antiquarks, the interaction rates are
increasing with higher temperature for all $\mu_q$. The charm quark
interaction rate with light quarks ($R_{uc}$) depends on
($T$,$\mu_q$) as described in Fig.
\ref{fig:cRatevsTmu} (a). For larger values of
$\mu_q$ and small temperatures ($T < T_c(\mu_q=0)$), $R_{uc}$ is much
larger than $R_{\bar{u}c}$ and $R_{gc}$, such that the total
interaction rates are dominated by $R_{uc}$. On the other hand the
$R$ profile is dominated by $R_{gc}$ for small $\mu_q$
and large temperatures. This is easy to interpret:
At large $\mu_q$ the number of light quarks is large compared to the
number of antiquarks, i.e. the Fermi-Dirac distribution contributes
differently to $R$ for $u$ and $\bar{u}$. On the other hand the gluon
number decreases with larger $\mu_q$ (relative to light quarks) since it is correlated with the
subdominant light antiquarks, via the $T$ and $\mu_q$ dependencies of the masses.

%----------------------------------------------------------------------------------------------------------------------------------------------------------------------------
\subsection{Diffusion coefficient and energy loss of charm quarks in
the hot medium} \label{HQDragDs}
%----------------------------------------------------------------------------------------------------------------------------------------------------------------------------
\vskip -2mm

Having  the matrix elements and the cross sections specified
we can calculate transport coefficients  $\mathcal{X}$ defined by \cite{Berrehrah:2014kba}
\begin{eqnarray}
\label{equ:Sec1.1}
&&\hspace*{-0.7cm}\frac{d <\!\!\mathcal{X}\!\!>}{d t} = \sum_{q,g} \frac{1}{(2 \pi)^5 2 E_Q} \int \frac{d^3 q}{2 E_q} f (\boldsymbol{q})
\int \frac{d^3 q'}{2 E_{q'}} \int \frac{d^3 p'_Q}{2 E'_{Q}}\nonumber \\
& & \hspace*{1.2cm} \times \ \delta^{(4)} (P_{in} - P_{fin}) \ \mathcal{X} \ {\frac{1}{g_Q g_p}}
|\mathcal{M}_{2,2}|^2\!,
%\nonumber\\
%& & {}
\end{eqnarray}
where $p'_Q$ ($ E'_Q$) is the final momentum (energy) of the heavy
quark with the initial energy  $E_Q$. In Eq. (\ref{equ:Sec1.1}) $q$ ($E_q$) and $q'$ ($
E_{q'}$) are the initial and final momenta (energies) of the partons
and $f(\boldsymbol{q})$  is their thermal distribution whereas
$|\mathcal{M}_{2,2}|^2$  is the transition matrix-element squared
for 2 $\to$ 2 scattering . Furthermore, in (\ref{equ:Sec1.1}) $g_Q$ is the
degeneracy factor of the heavy quark ($g_Q = 6$) and $g_p$ is the
degeneracy factor, i.e.  $g_p$=16 for gluons and $g_p$=6 for light
quarks. We mention that in Eq. (\ref{equ:Sec1.1}) we have
discarded Pauli blocking or Bose enhancement factors $(1 \pm f({\boldsymbol
p}')$ in the final states  since in our case the occupation numbers
$f({\boldsymbol p}')$ are rather small in the temperature range of interest
due to the rather massive degrees of freedom with pole masses larger
than twice the temperature. The errors introduced in this way are
smaller than the systematic errors incorporated in
$|\mathcal{M}_{2,2}|^2$, i.e.  in the transition matrix-element
squared. Employing $\mathcal{X}$ = $(E - E')$ we can calculate the
energy loss, $d< E>/d t (p_Q, T)$, whereas $\mathcal{X}$ =
$(\boldsymbol{p}_Q - \boldsymbol{p}'_Q)$ gives the drag coefficient,$d
<\boldsymbol{p}_Q>/d t  = A (p_Q, T)$.

The spatial diffusion coefficient $D_s$ can be expressed in two different ways \cite{Moore:2004tg}.
It can be obtained from the  slope of the drag coefficient divided by the heavy quark
momentum $\eta_D = A/p_Q$,
\begin{equation}
D_s =\lim_{p_Q\to 0} T/(M_Q \eta_D), \label{eq7} \end{equation} as
in Ref. \cite{Berrehrah:2014kba}. It can also be obtained from the
diffusion coefficient $\kappa = \frac{1}{3} d<(\boldsymbol{p}_Q -
\boldsymbol{p}'_Q)^2>/ dt$, calculated by Eq.  (\ref{equ:Sec1.1}), as in
Ref. \cite{Tolos:2013kva}
\begin{equation}
D_s = \lim_{p_Q\to 0} \frac{\kappa}{2 M_Q^2 \eta_D^2} . \label{eq9}
\end{equation} Both definitions agree if  the Einstein relation is
valid. Since in the case of the DpQCD  model the deviation
from the Einstein relation for small momenta $p_Q$ is of the order
10-15\% we will adopt eq.(\ref{eq9}) for the calculations shown here.
We note that the relation (\ref{eq7}) is strictly valid in the non-relativistic limit where bremsstrahlung is negligible,
i.e. for velocities $\gamma \ v < 1/\sqrt{\alpha_s}$ to leading logarithm in $T/m_D$, where $m_D$ is the Debye mass.
Therefore, it is a good approximation for the interaction of thermal heavy quarks, $M_Q \gg T$, with a
typical thermal momentum $p \sim \sqrt{M T}$ and a velocity $v \sim \sqrt{T/M} \ll 1$.
%\vspace{1truecm}
\begin{figure}[h!] %tbh!
\begin{center}
\includegraphics[scale=.21]{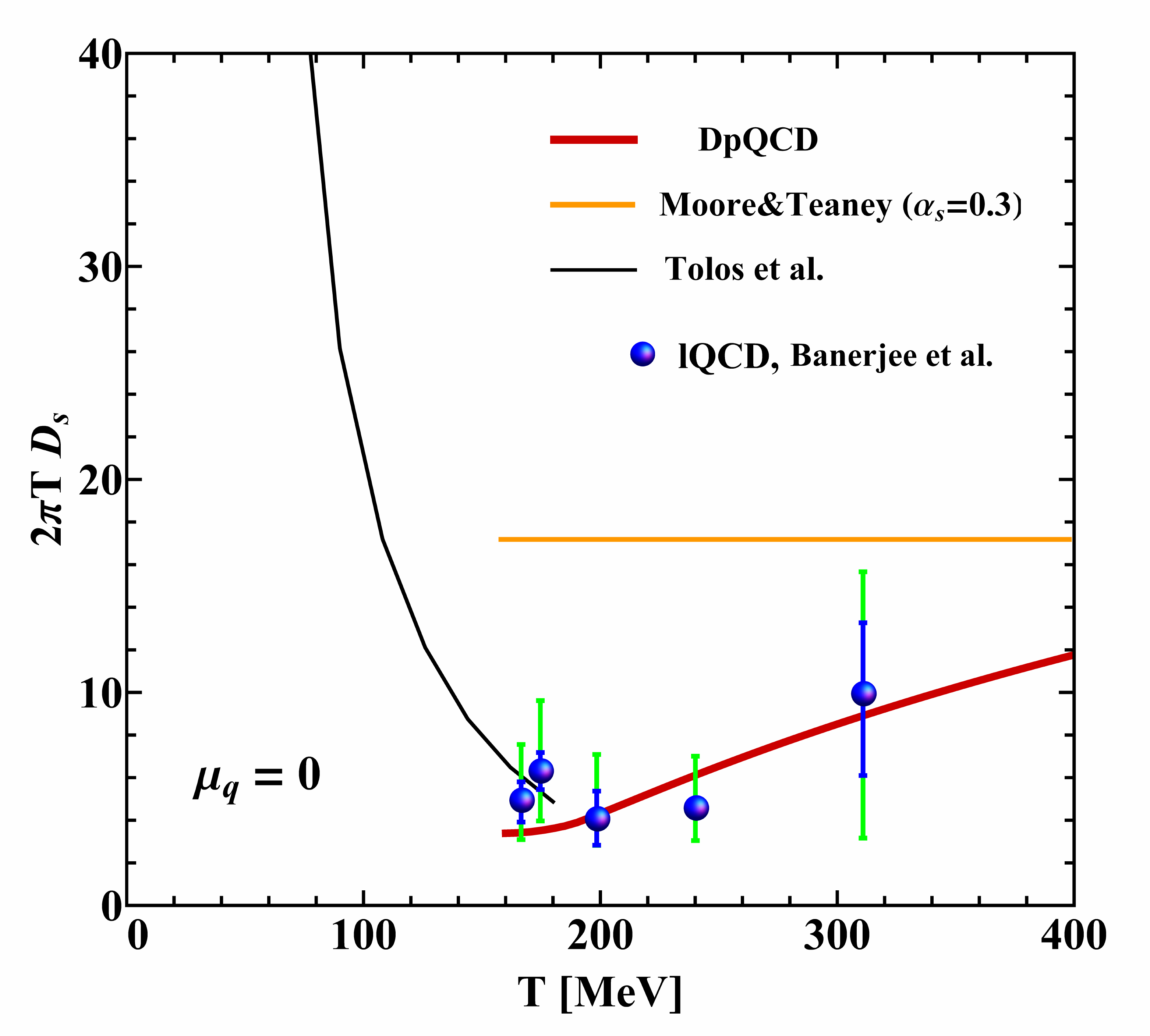}  \includegraphics[scale=.35]{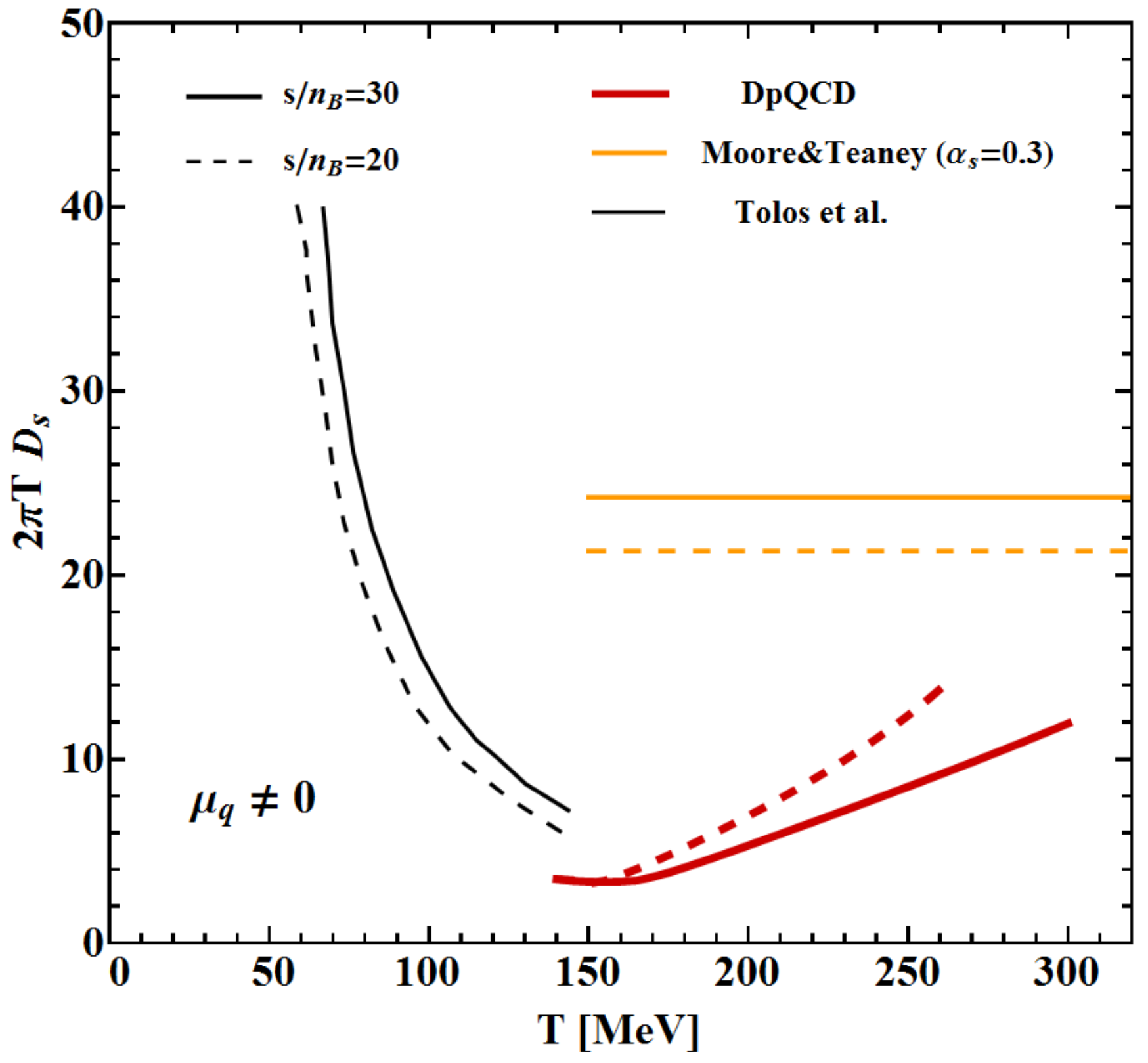}
\caption{\emph{(Color online)  (l.h.s.) Spatial diffusion coefficient for heavy quarks, $D_s$, as a function of
 $T$ for $\mu_q=0$.  Below $T=180\  MeV$ we display the hadronic diffusion
 coefficient  from
 Ref.\cite{Tolos:2013kva}, above $T=180$ MeV that for a partonic environment. The solid orange
 line is the result of Ref.\cite{Moore:2004tg} while the red thick solid line shows the DpQCD prediction.
 The lattice calculations are from Ref. \cite{Banerjee:2011ra}.}
 (r.h.s.) Spatial diffusion constant, $D_s$, as a function of $T$ for $\mu_q \neq 0$.
$D_s$ is displayed for different values of $s/n_B$ for a hadronic environment \cite{Tolos:2013kva}
as well as for a partonic environment. For the latter pQCD calculations are confronted with DpQCD
calculations. The figures are taken from Ref. \cite{Berrehrah:2014cc}. }
\label{fig:cdEdxTmu}
\end{center}
\end{figure}

In Fig. \ref{fig:cdEdxTmu} (l.h.s.) we display the spatial diffusion coefficient $D_s$ (\ref{eq9}) as a function
of $T$ for $\mu_q = 0$. Our results are compared with the leading
order (LO) results obtained by Moore and Teaney \cite{Moore:2004tg}
for perturbative partons and $\alpha_s $ = 0.3 as well as with the
lattice calculations from Ref. \cite{Banerjee:2011ra} for
temperatures above $T_c$.  We mention that a Debye mass is included
in the Moore and Teaney calculation \cite{Moore:2004tg} for the scattering of heavy
quarks on the QGP partons and leads to a finite cross section at
vanishing 4-momentum transfer. This Debye mass is generated
dynamically and in some sense can be considered as the analogue to
the DQPM pole masses. The lattice results in Fig. \ref{fig:cdEdxTmu} (a) have recently
been confirmed by the Bielefeld collaboration.

The spatial diffusion coefficient in deconfined matter is compared
with the result for the spatial diffusion coefficient of a heavy
meson in hadronic matter \cite{Tolos:2013kva} in
Fig.\ref{fig:cdEdxTmu} (l.h.s.) for temperatures below $T_c$.
 We observe that at $T \approx T_c$ the spatial diffusion coefficients for hadronic
and partonic matter join almost continuously and agree with the
lattice results. On the other hand, pQCD calculations yield a larger
value of the spatial diffusion coefficient as compared to the DpQCD
model leading to a discontinuity of $D_s$ close to $T_c$.
Rapp et al. \cite{Rapp:2009my,He:2012df} have shown that the spatial diffusion coefficient in
pQCD calculations can be lowered by adding nonperturbative heavy-quark interactions. Also hard thermal loop calculations
with 'effective' Debye masses and a running coupling lead to a
substantial lowering of $D_s$ and bring its values to the vicinity
of the lattice results  \cite{Berrehrah:2014kba}.

The DpQCD calculations can be extended to finite $\mu_q$ assuming
adiabatic trajectories (constant entropy per net baryon
$s/n_B$) for the expansion.  The latter is calculated
using the pole masses of the plasma constituents in the DQPM model
as well as for perturbative partons. For a given $s/n_B$ the
chemical potential $\mu_B$ is a monotonic function of $T$ and therefore we can display  $D_s$ as a function
 of $T$  and  $s/n_B$. Fig. \ref{fig:cdEdxTmu} (r.h.s.) displays the spatial diffusion coefficient for finite chemical
 potential, i.e for different values of the entropy per net baryon $s/n_B$. The pQCD calculations are
 obtained by adding the chemical potential to the thermal distributions and to the Debye mass when
 calculating the pQCD drag and diffusion coefficients (cf. Eq. (B13) of Ref.\cite{Moore:2004tg}).
 We observe - as in the $\mu_q = 0$ case - that the DpQCD spatial diffusion coefficient of
 heavy quarks approximately joins smoothly those of the hadron gas. (We expect that in the $\mu_B$
 region investigated here
the transition remains a cross over transition). On the contrary, pQCD
 calculations close to $T_c$ are a factor of 3 higher leading to a discontinuity of the spatial
 diffusion coefficient, which is not compatible with a cross-over transition as predicted by lattice calculations.
 This is a strong indication that close to the phase transition the effective degrees-of-freedom should be  massive
 quasi-particles and not perturbative quarks and gluons.

\begin{figure}[h!] %tbh!
\begin{center}
\begin{minipage}{0.45\textwidth}
\includegraphics[scale=.35]{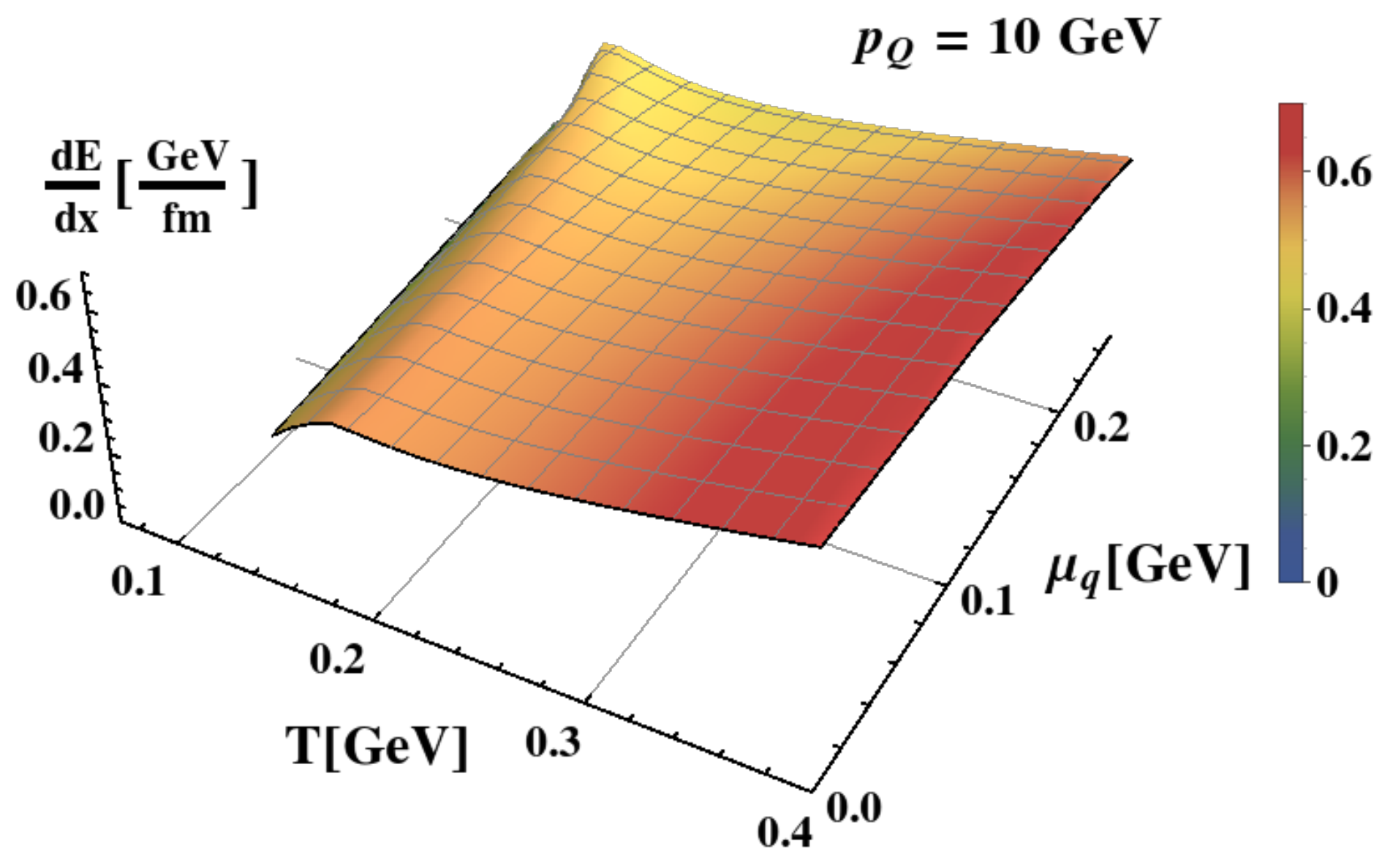}
\end{minipage} \hspace*{1.5cm}
\begin{minipage}{0.35\textwidth}
\caption{\emph{(Color online) Energy loss per unit length, $dE/dx$, of a c - quark with incoming momentum
of $10\ GeV/c$ in the plasma rest frame as a function of the temperature and quark chemical potential.
The figure is  taken from Ref. \cite{Berrehrah:2014cc}. }}
\label{fig:EdxTmu}
\end{minipage}
\end{center}
\end{figure}

For comparing our model predictions with experimental data another
transport coefficient, the energy loss of a heavy quark per unit
length, $d<E>/dx=  d<E>/vdt$, is important. It can
be obtained from Eq. (\ref{equ:Sec1.1}) by the choice $\mathcal{X}$
= $(E_Q - E'_Q)$. The energy loss  of a heavy quark with an incoming
momentum of  10 GeV/c as a function of $T$ and $\mu_q$  in the DpQCD
approach is presented in Fig. \ref{fig:EdxTmu}. As expected for a
cross-over transition we observe a very smooth dependence on both
variables, $T$ and $\mu_q$. For $\mu_q =0$ the gluon pole mass
depends on the temperature and therefore the increase of the energy
loss is due to the change of the running coupling
$g^2(T/T_c)$. For $\mu_q =0.2$ GeV, the energy loss is also
increasing with temperature but less than for $\mu_q=0 $ because
here the increase of the coupling is partially counterbalanced by
the decrease of the gluon pole mass.

We recall that the properties of the QCD medium in terms of the
shear viscosity over entropy ratio $\eta/s$ as well as the electric
conductivity over temperature $\sigma_e/T$ show a minimum close to
$T_c$ (cf. Section 4) which apparently
repeats in the charm spatial diffusion coefficient reflecting a
maximum in the interaction strength $g^2(T/T_c)$ of the QCD
degrees of freedom at temperatures close to $T_c$.

%--------------------------------------------------------------------------------------------------------------------------------------------------------------------------------
\section{Summary}
\label{summary}
%--------------------------------------------------------------------------------------------------------------------------------------------------------------------------------

We have presented in this review an extended dynamical quasiparticle
model (DQPM$^*$) incorporating  momentum-dependent selfenergies in
the parton propagators which are reflected in momentum-dependent
masses and widths. Accordingly, the QGP effective degrees of freedom
appear as interacting off-shell quasi-particles with masses and
widths that depend on three-momentum ${\boldsymbol p}$, temperature $T$ and
chemical potential $\mu_q$ as given in Eqs. (\ref{equ:Sec2.1}).
These expressions provide a proper high temperature limit (as in the
HTL approximation) and approach the pQCD limit for large momenta
$|{\boldsymbol p}|$. As in the standard DQPM the effective coupling is
enhanced in the region close to $T_c$, which leads to an increase of
the parton masses roughly below 1.2 $T_c$ (cf. Fig. 1 (a)).
%We
%mention that instead of displaying the parton masses as a function
%of temperature one may alternatively display them as a function of
%the scalar parton density $\rho_s(T)$ (cf. Ref.
%\cite{Bratkovskaya2011162}) and incorporate the quasiparticles in
%relativistic transport models.
%interpret the masses as a scalar
%mean-field depending on $\rho_s$.
%Since
%$\rho_s$ is a monotonically increasing function with temperature $T$ the
%masses $M_j(\rho_s)$  show a minimum in $\rho_s$ for $\rho_s
%\approx $ 0.5-1 fm$^{-3}$ which specifies the parton scalar density where the
%effective interaction -- defined by the derivative of the scalar mean field
%with respect to the scalar density -- changes sign, i.e. the net
%repulsive interaction at high scalar density becomes attractive at
%low scalar density and ultimately leads to color neutral bound states of the
%constituents (cf. Ref. \cite{Cassing2007365}).

The extended dynamical quasiparticle model DQPM$^*$ reproduces quite
well the lQCD results, i.e. the QGP equation of state, the baryon
density $n_B$ and the quark susceptibility $\chi_q$  at finite
temperature $T$ and quark chemical potential $\mu_q$  which had been
a challenge for quasiparticle models so far \cite{Plumari:2011mk}
(see also Fig. 4b). A detailed comparison between the available
lattice data and DQPM$^*$ results indicates a very good agreement
for temperatures above $\sim$ 1.2 $T_c$ in the pure partonic phase
and therefore validates our description of the QGP thermodynamic
properties. For temperatures in the vicinity of $T_c$ (and $\mu_B$=
400 MeV) we cannot expect our model to work so well since here
hadronic degrees of freedom, which are discarded in the DQPM$^*$,
mix in a crossover phase.

Furthermore, we have computed also the QGP shear viscosity $\eta$,
the bulk viscosity $\zeta$, and the electric conductivity $\sigma_e$
at finite temperature and chemical potential in order to probe some
transport properties of the medium. The relaxation times at finite
temperature and chemical potential, used in our study, are evaluated
for the dynamical quasi-particles using the parton width which is
averaged over the thermal ensemble at fixed $T$ and $\mu_q$. We,
furthermore, emphasize the importance of nonperturbative effects
near $T_c$ to achieve a small $\eta/s$ as supported by different
phenomenological studies and indirect experimental observations as
well as a maximum in the ratio $\zeta/s$. When comparing our results
for $\eta/s$ to those from the standard DQPM (with
momentum-independent selfenergies) in Ref. \cite{Ozvenchuk:2012kh}
we find a close agreement. In the DQPM$^*$ the gluon mass is
slightly higher (for low momenta)  and the quark mass is slightly
smaller than in the DQPM. Furthermore, the interaction widths are
somewhat larger in the DQPM$^*$ which finally leads to a slightly
lower shear viscosity $\eta$ than in the DQPM. This also holds for
the electric conductivity $\sigma_e$ which in the DQPM$^*$ gives
results even closer to the present lQCD 'data'.

Additionally, we have reported on the results of the
momentum-independent DQPM for the description of heavy quarks $Q$ in
the hot partonic medium and studied their transport properties, i.e.
the spatial diffusion coefficient $D_s$ and the energy loss $dE/dx$
also at finite $T$ and $\mu_q$. A medium at finite chemical
potential leads to a reduction of the $qQ$ and $gQ$ elastic cross
section and consequently to a reduction of heavy-quark energy and
momentum losses as compared to a medium at $\mu_q$=0. Nevertheless,
we have concluded that longitudinal momentum transfers are important
not only in a hot medium but also in a dense medium whereas the
dense medium leads to less transverse fluctuations in the heavy
quark propagation. The relative large drag at low temperatures  is
due to the strong increase of the running coupling $\alpha_s (T,
\mu_q)$ (infrared enhancement) for temperatures close to $T_c
(\mu_q)$.

Furthermore, we have observed a smooth dependence of the energy loss $dE/dx$ on both
variables $T$ and $\mu_q$ at finite but not too large values of
$\mu_q$. Such a profile is expected for a cross-over transition from
the partonic to the hadronic medium. For $\mu_q = 0$ the gluon mass
depends on the temperature and therefore the increase of the energy
loss is due to a change of the coupling. For $\mu_q = 0.2$ GeV, the
energy loss is also increasing with temperature but less than for
$\mu_q = 0$ because here both the coupling and the effective gluon
mass decrease and the increase of the infrared regulator is
counterbalanced by the decrease of the coupling. Since the
variations of all transport coefficients with $T$ and $\mu_q$ are
rather smooth (within the present DQPM/DQPM$^*$ propagators) the transition
from hadronic degrees of freedom to partonic ones remains a
crossover up to $\mu_q$ = 0.2 GeV.

In view of our results on the description of the QGP thermodynamics
and transport properties, one can conclude that the DQPM$^*$
provides a promising approach to study the QGP in equilibrium at
finite temperature $T$ and chemical potential $\mu_q$. Moreover, we
have demonstrated that one can simultaneously reproduce the lQCD
pressure, the quark susceptibility and the QCD transport properties
using a dynamical quasi-particle picture for the QGP effective
degrees of freedom that allows for a transparent interpretation of
the various results from lattice QCD.

We recall that a covariant transport approach has been set up a
couple of years ago in Refs.
\cite{Cassing:2009vt,Bratkovskaya2011162} in which the description
of the partonic phase has been based on the partonic propagators of
the standard DQPM. This approach is denoted by
parton-hadron-string-dynamics (PHSD) and has been employed for the
description of $p+p$, $p+A$ and $A+A$ reactions at invariant
energies from  $\sqrt{s_{NN}} \approx$ 8 GeV to 5 TeV. For a recent
review on bulk and electromagnetic probes we refer the reader to
Ref. \cite{Linnyk:2015rco} and for an application to the charm
sector at RHIC and LHC energies to Refs.
\cite{Song:2015ykw,Song2015sfa}. Since the DQPM$^*$ provides
appropriate propagators also for finite quark chemical potentials
$\mu_q$, a related implementation in the PHSD is foreseen and will
allow to investigate the phase boundary in heavy-ion collisions also
at lower bombarding energies (FAIR/NICA) where baryonic effects and
chiral symmetry restoration in the hadronic phase  are expected to
dominate \cite{CSR2016}.

%--------------------------------------------------------------------------------------------------------------------------------------------------------------------------------
\section*{Acknowledgements}
\label{acknowledgment}
%--------------------------------------------------------------------------------------------------------------------------------------------------------------------------------
The authors acknowledge valuable discussions with R. Marty, P. B.
Gossiaux, J. Aichelin, O. Linnyk, P. Moreau, A. Palmese, E. Seifert and T. Song. This
work has been supported by the ``HIC for FAIR'' framework of the
``LOEWE'' program. The computational resources have been provided by
the LOEWE-CSC.

\bibliographystyle{ws-ijmpe}
\bibliography{References}

\end{document}